\pdfoutput=1
\documentclass{article}
\usepackage[margin=1in]{geometry}

\usepackage{mystyle}
\usepackage{mynotation}

\title{AI Supply Chains: An Emerging Ecosystem of \\AI Actors, Products, and Services}

\author{
	Aspen Hopkins${}^{*,1}$, Sarah H. Cen${}^{*,2}$, Isabella Struckman${}^{1}$, \\ \vspace{-8pt}
	Andrew Ilyas${}^{3}$, Luis Videgaray${}^{4}$, Aleksander M\k{a}dry${}^{1}$
}
\date{}

\begin{document}
	
	\maketitle
	
	\renewcommand{\thefootnote}{\fnsymbol{footnote}}
	\footnotetext[1]{Equal contribution. Correspondence to \texttt{dataspen@mit.edu} and \texttt{shcen@stanford.edu}}
	
	\renewcommand{\thefootnote}{\arabic{footnote}}
	\footnotetext[1]{Department of Electrical Engineering and Computer Science, Massachusetts Institute of Technology}
	\footnotetext[2]{Department of Computer Science and Stanford Law School, Stanford University}
	\footnotetext[3]{Department of Statistics, Stanford University}
	\footnotetext[4]{Sloan School of Management, Massachusetts Institute of Technology}
	\setcounter{footnote}{4}

	\begin{abstract}
		The widespread adoption of AI in recent years has led to the emergence of AI supply chains:
complex networks of AI actors contributing models, datasets, and more to the
development of AI products and services. 
AI supply chains have many implications yet are poorly understood. 
In this work, we take a first step toward a formal study of AI supply chains and their implications, 
providing two illustrative case studies indicating that both AI development and regulation 
are complicated in the presence of supply chains. 
We begin by presenting a brief historical perspective on AI supply chains,
discussing how their rise reflects a longstanding shift towards specialization and outsourcing that signals the healthy growth of the AI industry.
We then model AI supply chains as directed graphs and demonstrate the power of this abstraction 
by connecting examples of AI issues to graph properties. 
Finally, we examine two case studies in detail, providing theoretical and empirical results in both. 
In the first, we show that information passing (specifically, of explanations) along the AI supply chains is imperfect, 
which can result in misunderstandings that have real-world implications. 
In the second, we show that upstream design choices (e.g., by base model providers) 
have downstream consequences (e.g., on AI products fine-tuned on the base model).
Together, our findings motivate further study of AI supply chains and their 
increasingly salient social, economic, regulatory, and technical implications. 

	\end{abstract}

\section{Introduction}
\label{sec:intro}

Most modern AI systems are no longer developed in-house, at least not in their entirety.
Instead, multiple organizations contribute to their development by providing resources, models, and datasets. 
This phenomenon has led to the rise of \emph{AI supply chains}: complex networks of AI components (and the respective actors who provide them) that contribute to the production of AI systems.\footnote{Note that we are differentiating ``AI \emph{in} supply chains'', 
which refers to the popular use of AI to optimize traditional supply chain processes, 
from ``AI supply chains'', through which ML systems and services are developed and distributed.}

\begin{figure}
    \centering
    \hspace*{-4mm}
    \includegraphics[width=.65\columnwidth]{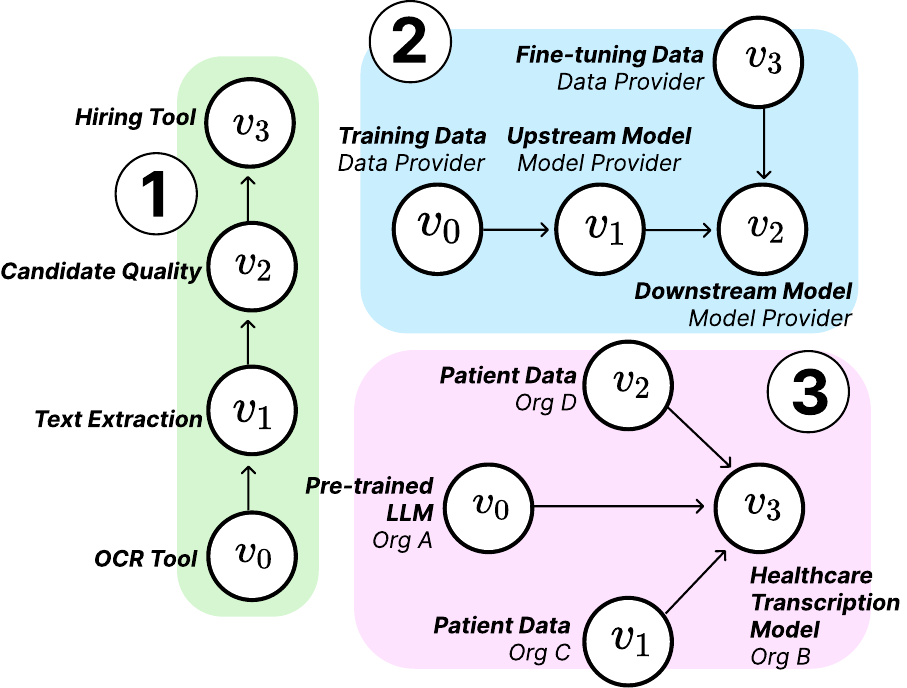}
    \caption{
    Three examples of AI supply chains. 
    (1) A linear AI supply chain where an OCR tool ($v_0$) contributes to a text extraction tool ($v_1$) used when predicting candidate quality ($v_2$), terminating in a hiring tool ($v_3$).
    (2) A generic upstream model ($v_1$) trained on some dataset ($v_0$) that is fine-tuned ($v_2$) using a new dataset ($v_3$).
    (3) A healthcare transcription model ($v_3$) that is composed of one pre-trained LLM ($v_0$) and patient data from two hospitals ($v_1$, $v_2$).
    }
    \label{fig:examples}
\end{figure}

To illustrate, 
consider an AI healthcare app that transcribes and summarizes doctors' visits, 
as visualized in Figure \ref{fig:examples}. 
Today, such a product is typically built via the efforts of several organizations.
One organization
(\texttt{Org A})
may develop a pre-trained large language model (LLM), while another---the healthcare app developer (\texttt{Org B})---may fine-tune it. 
In fine-tuning, suppose \texttt{Org B} then uses datasets curated by two unrelated hospitals (\texttt{Org C} and \texttt{Org D}). 
The result is an AI product built from the contributions 
of \emph{at least} four participants in the AI supply chain, though numerous others may contribute additional resources such as compute, upstream datasets, and services. 

The rise of AI supply chains reflects a change in the status quo of AI development that was accelerated by
generative AI and subsequent general-purpose models \cite{lee2023talkin}, 
such as OpenAI's GPT
and Meta's Llama. 
Although early AI supply chains can be traced back several decades (e.g., to the outsourcing of data labeling and curation), 
general-purpose models sparked an explosion of AI adoption
by making it accessible and affordable. 
Indeed, in the few years since the popularization of general-purpose models,
between 35 and 67 percent of businesses now report integrating AI into their operations \cite{springsapps2024,ah2024}, 
with more and more companies rushing to meet the growing demand for AI.
The resulting ecosystem of AI actors and components is what we refer to as AI supply chains. 
As the AI industry matures, 
supply chains---which lower costs in development and allow for specialization---will be key to AI's proliferation.

The rapid growth of AI supply chains and their ability to facilitate AI adoption introduces new challenges and risks to AI development, safety, and regulation.
By distributing AI development across multiple actors, AI supply chains blur the responsibilities held by stakeholders, 
complicating efforts to enforce quality controls or assign liability.
Poor or non-existent communication between organizations creates opportunities for failures and disruptions 
(e.g., a flaw in GPT's model update)
to cascade down the AI supply chain and affect countless downstream components.
Moreover, the disparate contributions of multiple organizations to a single AI product hinders regulatory efforts to monitor AI systems and perform end-to-end auditing.
On top of all this, the growth of AI supply chains may concentrate market power in the hands of a few companies that sit at the top, 
allowing them to take unilateral actions that affect the entire supply chain.

Several other works have begun to document AI supply chains and their impacts,
including \citet{lee2023talkin} who explore their implications for copyright; 
\citet{widder2023dislocated} and \citet{cobbe2023understanding} who investigate their effects on AI accountability; 
and \citet{aisupplychains23} who examine the potential economic and policy implications. These works establish that AI supply chains carry social, economic, and political importance.
Our work adds to this body of research, 
focusing on how \emph{machine learning} outcomes change with the emergence of AI supply chains. To our knowledge, this work is the first to formalize the behavior of machine learning systems developed across AI supply chains.

\paragraph{Our contributions.}
Our main contributions are to introduce AI supply chains through a formal analysis and to study the machine learning challenges that arise from spreading AI development across them.
In Section \ref{sec:aisc_setup}, we begin by discussing how AI supply chains emerged and how their recent growth indicates that the AI industry is maturing.
In Section \ref{sec:SCNs}, we present a model of AI supply chains as directed graphs, 
where nodes correspond to AI components (and their corresponding AI actors). 
We illustrate the usefulness of this model by connecting issues in the AI supply chain to graph properties.

In Sections \ref{sec:explanations} and \ref{sec:fairness},
we present two case studies that demonstrate how AI supply chains affect machine learning outcomes. 
In both case studies, we provide formal results alongside experiments.
The first case study examines how information propagation along the AI supply chain can be imperfect.
Motivated by a setting in which there are explainability requirements (e.g., due to regulation), 
we show that explanation fidelity degrades with the length and width of the supply chain, 
suggesting that explanation requirements must be reassessed when an AI decision is the result of an AI supply chain. 
The second case study examines how upstream design choices affect downstream outcomes. 
To do so, we focus on design choices related to algorithmic fairness. 
We find that when upstream models satisfy certain notions of algorithmic fairness,
downstream models experience different performance-fairness trade-offs, even if the downstream notion 
of fairness is different from the upstream one.\footnote{We note that our findings do not indicate that algorithmic fairness is a restrictive criterion that should not be sought. Rather, our results suggest that all design choices (even the choice to not pursue fairness upstream) have downstream consequences.}
Together, our analysis is intended to inform AI development and evaluation as AI supply chains continue to grow.

		\section{Background and Related Work}
\label{sec:aisc_setup}

In this section, we discuss the emergence of supply chains in AI development. 
We begin with a brief historical perspective, 
then examine related work on AI supply chains.

\subsection{Rise of AI Supply Chains}

Early machine learning (ML) development took place in silos.
Single organizations (and often individual researchers)
would build a predictive model end-to-end, from the data collection and model selection to the training and evaluation. 
With few exceptions, organizations did not outsource ML development.

The last two decades have seen ML (and subsequently AI) pipelines fragment at a rapid pace.
Early examples of this fragmentation occurred in \emph{datawork}.
Curating datasets is a costly endeavor that requires collecting, labeling, and cleaning thousands (if not millions) of samples. 
One major step that the ML community took was the release of open datasets such as WordNet \cite{wordnet}, MNIST \cite{6296535}, and ImageNet \cite{5206848}, 
which enabled ML researchers to train on existing datasets rather than undertake the expensive operation curating them. 
These datasets (alongside dataset libraries, 
including UCI Machine Learning Repository \cite{frank2010uci} and Kaggle \cite{kaggle}) accelerated ML development and set the scene for later commercialization of datawork by services, such as Mechanical Turk for data labeling.

There were analogous developments around the outsourcing of \emph{model training}.
As ML models grew, researchers sought ways to achieve high performance with fewer resources (e.g., less compute). 
One concept that arose was transfer learning \cite{pan2009survey,caruana1994learning, bengio2012deep}: a family of methods for adapting existing models to new tasks that encompasses, e.g., fine-tuning \cite{yosinski2014transferable}.
This approach allowed researchers to achieve state-of-the-art results without training models from scratch. 
Outsourcing all or part(s) of training became popular through services such as APIs, fine-tuning or post-training services, and AutoML tools \cite{hutter2019automated}.

This shift toward outsourcing AI development rapidly accelerated after 2022. 
Generative AI gave rise to a wave of AI models, 
such as OpenAI's GPT, Anthropic's Claude, and Google's Gemini.
Each became ``base'' models that businesses could repurpose and adapt for their own use, 
earning their namesake as ``general-purpose'' models (also known as ``foundation'' models \cite{bommasani2021opportunities}). 
By lowering barriers to entry in AI and resetting expectations for previously cost-prohibitive performance, 
generative AI and subsequent general-purpose models incentivized organizations to accept upstream dependencies.
Stated simply, general-purpose models made AI accessible by saving businesses from having to train powerful models from scratch.
As such, many (though not all) modern-day AI supply chains follow a similar recipe: 
a large dataset as the ``raw material,''
a base model that is produced by training on this dataset, 
and a domain-specific tool (e.g., a doctor-patient visit summarization app) that is produced by fine-tuning the pre-trained base model on a specialized (e.g., hospital) dataset. 
Around general-purpose AI, 
there is now an array of actors rising to meet the demand for AI. 
Companies such as Microsoft, Amazon, Google, Scale AI, and Databricks offer suites of AI product and services,
and there is a growing industry around data centers to support the massive compute needs of AI.
We anticipate that the AI supply chain will continue to expand, thus motivating further study.

Before proceeding, 
we note that this fragmentation into supply chains is not inherently harmful.
In fact, it as a natural
shift that signals the AI industry's healthy growth. Supply chains allow tasks to be divided and outsourced, 
which not only reduces costs, but also allows organizations to specialize in a specific product or service.
As such, the emergence of AI supply chains 
indicates that the AI industry is taking steps toward efficiency and specialization.

\subsection{Related Work}

The rising prevalence of AI supply chains has brought with it a growing academic interest in their effects. 
For example, \citet{widder2023dislocated} and \citet{cobbe2023understanding} examine 
the social and ethical considerations of having multiple external contributors to an
ML product. \citet{lee2023talkin} discuss the legal implications, particularly with regard to copyright violation, 
while \citet{attard2023valuing} consider Canada's efforts in regulating AI through AI supply (or value) chains. 
\citet{bommasani2023ecosystem} documents examples of AI supply chains in the wild 
via ``ecosystem graphs,'' with an emphasis on the role of large pre-trained 
foundation models.  
In considering the impacts of foundation models,
\citet{suresh2024participation} explore how participation can shape
general-purpose AI models by targeting downstream applications. 
Our work contributes to this growing area of research, 
with an emphasis on their technical implications.

Although not explicitly studied in the context of AI supply chains,
there is a mature body of work in ML on problems that we can view as constituent 
pieces of AI supply chains. 
For example, transfer learning \cite{krizhevsky2012imagenet}, and rapidly growing contributions to modular multi-modal models \cite{xu2023mplug} show how model components may be adapted to new settings.
A separate line of work in algorithmic fairness \cite{dwork2018fairness} studies 
the problem of {\em fairness under composition}, 
where the goal is to reason about the behavior of
distinct ML models.
For data, work in federated learning \citet{mcmahan2017communication},
data poisoning \cite{biggio2012poisoning},
and backdoor attacks \cite{khaddaj2023rethinking} explore how ``upstream'' AI decisions affect (and in particular, break) downstream expectations. 
Each of these directions (amongst numerous others) contributes to our technical understanding of the challenges faced in fragmented AI development.

		\section{AI Supply Chains As Directed Graphs}
\label{sec:SCNs}

Despite the relative nascence of the AI industry,
AI supply chains are remarkably intricate.
They consist of models (pre-trained, fine-tuned,
or post-trained),
datasets (raw or curated, labeled or unlabeled),
predictions,
benchmark evaluations, AI services,
and more. 
We refer to the above as ``components.''
Moreover, the actions that AI actors can take on these components are varied and complex. 
For example, given a model as the ``component'' of interest, 
an AI actor can take multiple actions: 
(i) they can \textit{adapt} a pre-trained model to a specific context \cite{hu2021lora,zhuang2020comprehensive};
(ii) they can \textit{aggregate or synthesize} multiple models, as is done in certain ensembling methods like weight-averaging \cite{ganaie2022ensemble};
and (iii) they can \textit{train or predict using the outputs} of models (e.g., AI-generated scores or synthetic data). 
This is certainly not a comprehensive list (retrieval-augmented generation, or RAG, \cite{lewis2020retrieval} falls into a new category, as do cases where training or evaluation artifacts, e.g., learning gradients, might be shared between participants), 
but it illustrates the multitude of common actions that can occur on a single AI component.

Below, we propose directed graphs as a natural way to flexibly model AI supply chains and capture these interactions, 
where nodes (or vertices) correspond to AI components and links (or edges) indicate the dependencies between these components.

\subsection{Model}

Let an AI supply chain be given by a graph $G = (V, E)$, 
where each node $v \in V$ denotes an {\em AI component} (such as a model or dataset), 
and $(v_j, v_i) \in E \subset V \times V$  if and only if there is a directed edge from node $v_j$ to $v_i$.
A node $v_k$ is a \emph{parent} of $v_i$ if there exists a directed edge $(v_k, v_i) \in E$ from $v_k$ to $v_i$. 
Thus, an edge corresponds to an {\em operation} used to obtain a downstream component from its parent. 
A node $v_k$ is an \emph{ancestor} of $v_i$ (and $v_i$ is a \emph{descendant} of $v_k$) if there exists a directed path from $v_k$ to $v_i$,
i.e., there exists a path $(v_k, v_{k'}), (v_{k'}, v_{k''}), \hdots, (v_{k'''}, v_i) \in E$.

\begin{example}
Consider the example visualized in Figure \ref{fig:examples}.2. 
    Vertices $v_0$ through $v_3$ denote the AI components. 
    The  AI components  $v_1$ and $v_3$ are the parents of $v_2$ while
     $v_0$, $v_1$, and $v_3$ are ancestors of $v_2$. 
     Both $v_0$ and $v_3$ have no parents or ancestors. 
     The AI component $v_1$ has one parent, $v_0$, which is also its only ancestor. 
    The set of ancestors for $v_2$ is $\{ v_0, v_1, v_3 \}$.
\end{example}
In this work, 
we restrict our attention to components that correspond to models or datasets. 
We leave analysis of additional components 
(e.g., compute) 
to future work.  
To distinguish between organizations in our figures, we label vertices with their AI component \textit{and} the organization that owns it, e.g. (\texttt{GPT-4, OpenAI}). In cases where all the nodes are owned by a single organization, we omit the second label.\footnote{Organizations can encompass multiple autonomous or semi-autonomous divisions \cite{chandler1969strategy, williamson1975markets}, and their distinction and definition entails a large body of work. Our labeling is intended to account for this variation.}
 
\subsection{Examples}

Viewing AI supply chains as directed graphs helps formalize our conceptual understanding of them. 
To illustrate, we use the remainder of this section to discuss four issues that arise in AI supply chains, instantiating examples using the setup described above. 
In this section, let $h_v$ denote the mapping from $v$'s parents to $v$ for all $v \in V$, e.g., $v_2 = h_{v_2}(v_1, v_3)$ and $v_1 = h_{v_1}(v_0)$ in Figure \ref{fig:examples}.1.
    
\subsubsection{AI components do not mix modularly}\label{sec:non-modularity}
When components are modular, connecting or disconnecting them does not change their individual attributes---much like connecting Lego bricks does not change their shape or color. 
This modularity is beneficial in supply chains \cite{jayaram2018role}.
For example, modularity supports efforts to trace failures or assign liability, 
or to enforce IP licensing and distribute royalties. 
Modularity in AI supply chains can therefore be desirable.

However, AI supply chains tend to mix the contributions of AI components in a \textit{non-modular} way, much like a blended soup. 
Consider the transfer learning example in Figure \ref{fig:examples}.2, where model $v_1$ is adapted (or fine-tuned) to produce a new model $v_2$. 
Given $v_2$, one cannot easily discern the AI components that produced it.
In contrast to modular supply chain components, it is difficult to disentangle the effects of $v_2$'s ancestors on the final AI component.
We might therefore study non-modularity as follows.
\begin{example}
    One can think of non-modularity as arising when, 
	for a fixed AI supply chain $G$ and node $v$, 
	there is a non-unique mapping from $v$'s ancestors to $v$. 
	Formally, let $a_1, a_2, \hdots$ denote $v$'s ancestors and recall the definition of $h_v$.
	Then, 
	non-modularity is captured when there exist $(h_v, h_{a_1}, h_{a_2}, \hdots) \neq (h'_v, h'_{a_1}, h'_{a_2}, \hdots)$ such that the behavior of the resulting AI component $v$ is the same under either set of mappings. Intuitively, given the final AI component $v$, one cannot separate the components that gave rise to it in a modular way. 
\end{example}

\vspace{-3mm}
\subsubsection{AI supply chains exacerbate  transparency issues in AI} 
\label{sec:hidden_interactions}

As AI supply chains grow, 
there may be ``hidden interactions'' between components. 
More precisely, 
there may be interactions between the ancestors of an AI component $v$ that are difficult to identify from the perspective of $v$.
To make this concrete, 
observe that 
$v$ is typically aware of its parents: the AI components that directly contribute to $v$. 
In some cases, $v$ may even be aware of its grandparent nodes. 
However,
$v$ generally does not have full knowledge of the graph and may be unaware of interactions between its ancestors.

The longer or more complex an AI supply chain, the more likely it is for hidden interactions to present.
For instance, 
consider Figure \ref{fig:examples}.2. 
Suppose that $v_2$ does not have knowledge of $G$ beyond its parents. 
Now suppose that there is an edge from $v_0$ to $v_3$
that $v_2$ is not aware of. 
A number of consequences may ensue from this scenario. For example, the model may exhibit overconfidence on subsets of data included in both $v_0$ and $v_1$, leading to downstream misuse. Further, as the interaction was hidden, $v_2$ may not know to monitor for this possibility, and may similarly lack a mechanism for repair. Hidden interactions can thus introduce vulnerabilities by obscuring characteristics about AI components that may seem innocuous when considered independently, but disproportionately amplify harms when combined. Such interactions are the result of incomplete information in or regarding the AI supply chain and its components. Under our model, we might capture this phenomenon as follows.
\begin{example}
    Consider an AI component $v$. 
    Suppose that $v$ has knowledge of its ancestors, but only up to $m$ hops away. 
    That is, $v$ is aware of any ancestors that are within $m$ hops of $v$ and of the edges between these nodes. 
    Then, one can define hidden interactions as the interactions between ancestors further than $m$ hops away from $v$. 
    There may, for example, be a cycle that includes $v$ but has length $> m$. 
\end{example}

\subsubsection{Spreading control over multiple organizations compromises AI resilience} 
\label{sec:dispersed_control}

Another consequence of distributing parts of machine learning development across an AI supply chain is that control over AI development is ``dispersed'' across multiple organizations. 
Stated differently, 
a single organization does not, in general, 
have full control of the pipeline that generates its AI components. 
Dispersed control has ramifications on the robustness and resilience of AI products and services. 
If, for instance, an upstream model is updated, it may break downstream AI components or cause them to behave unexpectedly. 
When all AI components are controlled by the same organization, 
the teams developing each component can address the problem internally, 
avoiding communication and information-sharing barriers that exist across organizations. 
Perhaps more worryingly, 
actions taken upstream may not be reversible downstream;
as such, a downstream organization may not be able to %
achieve their own design objectives without changing their upstream AI vendors. 
Consider the following formalization. 

\begin{example}
    Suppose that every AI component $v \in V$ is associated with a different organization. 
    Suppose further that we consider $m$-dispersed control to exist for an AI component $v$ when 
    changes to ancestors within $m$ hops of $v$ cannot be ``reversed'' by any changes made to $h_v$.
    Formally,
    recall the definition of $h_w$ from Section \ref{sec:SCNs}.    Suppose that $h_w \in \mathcal{H}_w$ for model classes $\{ \mathcal{H}_w \}_{w \in V}$.
    Let $\mathcal{A}_m \subset V \setminus \{ v \}$ denote the ancestors of $v$ that are within $m$ hops of $v$.
    Then, $m$-dispersed control occurs at $v$ when there exists $ \{ h'_a \in \mathcal{H}_a \}_{a \in \mathcal{A}_m}$ such that no $h_v'$ is able to match the behavior $v$ under the original mappings $(h_v, h_{a_1}, h_{a_2}, \hdots)$.
\end{example}

\subsubsection{Cycles in AI supply chains facilitate feedback loops}
\label{sec:cycles}

As is often a concern in directed graphs, AI supply chains may contain cycles. 
For example, an LLM may be used by a downstream organization to generate articles that are then scraped as new training data for an updated version of the LLM. 
Many of the dangers caused by cycles in AI supply chains are amplified when there are hidden interactions (cf.  Section  \ref{sec:hidden_interactions})
or dispersed control (cf. Section \ref{sec:dispersed_control}). 
However, even in their absence, cycles present challenges to AI performance, fairness, and more.
Cycles in AI supply chains can be cast as the feedback loops previously discussed in machine learning, 
including \citet{ensignrunaway}'s work on feedback loops in policing, 
\citet{kleinberg2021algorithmic,bommasani2022picking}'s on algorithmic monocultures, 
and may hold parallels to homogenization mechanisms in recommendation systems 
\cite{chaney2018algorithmic}. Consider the following example.
\begin{example}
    Let $v_0$ be an LLM. Suppose that there are a set of downstream models $\{v_1, v_2, \hdots , v_m \}$ that generate text (such as articles, blogs, and resumes) and are descendants of $v_0$. Then, if $v_0$ is continually retrained on textual data obtained from these downstream models, i.e., there is a cycle from $v_0$ to $v_i$ and back to $v_0$ for all $i \in [m]$.
    Such cycles can create a feedback loop and lead to text homogenization. 
    It is possible that the effects of a feedback loop are dampened by, for instance, the existence of many incoming edges to nodes in the cycle that disrupt the feedback loop. 
    One potential method of measuring would be to compare, e.g., the betweenness centrality of nodes in and adjacent to the cycle. 
\end{example}

		\section{Case Study 1: Passing Information Along the AI Supply Chain}\label{sec:explanations}

In an AI supply chain, 
multiple actors have control over different parts of an AI pipeline. 
Since no single actor has full control over the pipeline, 
the actors coordinate by communicating information to one another.
In this section, 
we explore the implications of passing information along the AI supply chain.
Specifically, 
we show that \textbf{AI supply chains hamper the ability of downstream developers to provide accurate explanations of their predictions}. 
We motivate this problem by considering a setting in which a downstream developer is required to provide 
an explanation for each of their predictions.
However, the developer has limited access to upstream models and cannot probe them in the same 
way they would an in-house model to provide an explanation. 
The downstream developer must therefore produce an explanation that is \emph{built on explanations that the intermediate model providers produce}.
Below, we illustrate how downstream explanations built in this way can become increasingly fraught.

\subsection{Problem Setup}\label{sec:explanation_setup}

Consider a supply chain corresponding to an AI-powered hiring tool: given a 
candidate's information, the hiring tool makes a suggestion about whether
they should be hired.
That is, given an {individual} represented by a feature vector $\bx \in \bbR^{\rho}$,
the tool makes {\em a prediction} 
$y \in \bbR$ 
about their success as an employee.
Rather than develop the entire model in-house, the developer of this tool
leverages  a supply chain of third-party services to, for example, process the candidates' resumes and recommendation letters, 
thus forming an AI supply chain (an example of which can seen in Figure \ref{fig:examples}.1). 
Formally, 
one can express the downstream hiring tool
$f_v: \bbR^\rho \rightarrow \bbR$ as a composite function:  
$$f_v(\bx) = h_v (\bx, f_{p_1}(\bx), \hdots, f_{p_n}(\bx) ),$$ 
where $p_1, \hdots, p_n \in V$ are $v$'s parents.
Note that $f_v$ and $h_v$ are not the same, as $f_v$ acts on $\bx$, whereas $h_v$ acts on inputs that may themselves depend on $\bx$.

In this section, we characterize how AI supply chains affect the ability to provide accurate explanations. 
For simplicity we will focus on {\em locally linear explanations}, 
which includes methods such as LIME \cite{ribeiro2016should}.
\begin{definition}[Local linear explanation]
\label{def:locally_linear}
    A $\delta$-local linear explanation at $\bz$ for a model $g$ is a matrix $E_\delta(g, \bz)$
    satisfying
    \[
    E_\delta(g, \bz) \in \arg\min_{W} \mathbb{E}_{\mathbf{u}} ||
        g(\bz + \delta \bm{u}) - g(\bz)
        - W^\top \mathbf{u}
    ||_2^2, 
    \]
    where $\mathbf{u}$ are drawn uniformly at random from the unit ball in $\bbR^\rho$.
    That is, $E_\delta(g, \bz)$ is the best {\em linear}
    approximation of $g$ around the point $\bz$.
\end{definition}   
A locally linear explanation is a linear approximation of how $g$ behaves in the neighborhood of $\bz$, 
where the size of the neighborhood is controlled by $\delta$.
To see this, note that as $\delta \rightarrow 0$, 
$E_\delta(g, \mathbf{z})$ is equivalent to the Jacobian of $f$ at $\mathbf{z}$ if $g$ is differentiable.
More explicitly, $W$ as a linear mapping applied to perturbation $\mathbf{u}$;
    since $E_\delta(g, \mathbf{z})$ minimizes the right-hand side, 
    it is the choice of $W$ that (in expectation) best approximates the behavior of $g$ between $\bz$ and $\bz + \delta \mathbf{u}$
    for all choices of $\mathbf{u}$ in the $\delta$-ball around $\bz$.

    Although locally linear explanations are imperfect, 
they are often used in practice because they provide a first-order approximation of how small changes in $\bz$ affect $g$'s output.
We adopt this formulation as it includes popular approaches, such as LIME \cite{ribeiro2016should} and SHAP \cite{lundberg2017unified}.

\paragraph{Explanation error.}
In this section, we examine the scenario where the downstream developer is required to report an explanation of its model decision. 
Because the downstream developer can only access its own model and not upstream ones, 
it must construct its explanation based on its own model and the information communicated by its parents $p_i$;
namely, its parents' explanations of the parent models.

We then show that the error in the communicated explanations from a node's parents
can propagate the error in upstream explanations, 
leading to compounding error as the supply chain grows.
To gain intuition for why error can compound, 
note that it is generally infeasible to compute ${E}_{\delta}(g, \bz)$ exactly for $\delta > 0$, 
so organizations typically compute  empirical approximations $\hat{E}_{\delta}(g, \bz)$ instead (e.g., via linear regression). 
Fitting a locally linear explanation can take hundreds or even thousands of queries to each model of interest, 
and the explanations that developers provide thus always contain some approximation error.

\paragraph{Computing explanations.}
In this section, we will repeatedly make use of the ``chain rule''.
That is, when a developer (i.e., node) $v$ receives an explanation $\hat{E}_\delta(f_p, \bx)$ from each parent $p$, 
they combine their parents' explanations with the explanation of their 
own model $\hat{E}_\delta(h_v, \bz)$ to produce the \emph{full} explanation $f_v(\bx)$
using the chain rule.  
(Here, $\bz$ is the input that node $v$ receives, i.e., $\bz = ( \bx, f_{p_1}(\bx), \hdots, f_{p_n}(\bx) $.)

Formally,
given a node $i$ and its parents $\text{pa}(i)$,
then the explanation that $i$ computes given the explanation of its parents is given by:
\begin{align*}
    \hat{E}_\delta(f_i, \bx) 
    &= 
    \sum_{j \in \text{pa}(i)} 
    \hat{E}_\delta(h_i, \bz_i)^\top 
    \hat{E}_\delta(f_j, \bx)
\end{align*}
The chain rule is the appropriate way to combine locally linear explanations because 
$E_\delta(g, \bz)$ approaches the Jacobian of $g$ at $\bz$ (if $g$ is differentiable and locally smooth) as $\delta \rightarrow 0$.

\subsection{Explainability Across Supply Chains is Unreliable}

Suppose that an employer must explain why an application $\bx$ receives a decision $y$, but $\bx$ is first pre-processed by upstream services $f_{p_i}$.
How faithful is the employer's explanation to the actual AI pipeline's behavior?
In this section, we formalize the phenomenon that communicating across an AI supply chain 
can lead to downstream information being misleading. 
In particular, we show that, even for a linear supply chain, 
the ``error'' of a locally linear explanation is compounded by a constant
that grows exponentially in the depth of the supply chain. 
For convenience, 
we call the ``end-to-end'' explanation the explanation that is computed when a single actor
has access to the entire AI system and can compute the explanation directly, 
and we call the ``supply-chain'' explanation the explanation that is computed 
by combining the explanations provided by actors in the supply chain. 

	\begin{theorem}\label{prop:explanation}
		Consider the setup in \Cref{sec:explanation_setup}.
		Consider a node $1$ that is the descendant of a linear supply chain of depth $d$, 
		where $j + 1$ is the parent of node $j$. 
		Let $\hat{E}_\delta(h_j, \mathbf{z}_j) = {E}_\delta(h_j, \mathbf{z}_j) + \Delta_j $ for all $j$,
		where we assume each $\Delta_j$ is a random matrix with mean-zero entries drawn i.i.d. from a distribution with variance $\sigma^2$.  
		We assume that:
		(i) $\| E_\delta(h_i,\mathbf{z}_i) \|_2^2 \leq C_1$ 
		and $\| E_\delta(f_{i+1},\mathbf{x}) \|_F^2 \leq C_2$ for universal constants $C_1$ and $C_2$;
		(ii) $f_j$ and $h_j$ are locally smooth and differentiable at $\bx$ and $\bz_j$,
		respectively, for all $j$ and all $\bx$ of interest;
		(iii) $\Delta_j$ is independent of any quantities upstream of it;
		(iv) $\mathbb{E}[\bz_i] = [0]$ for all $i$;
		and 
		(v) $\text{dim}_2(\Delta_i) = \text{dim}_2(\Delta_j)$ for all $i, j \in [d]$.
		Note that (v) is not necessary but simplifies the expression.
		Then, 
		\begin{align*}
			\mathbb{E} \left[ \|\hat{E}_\delta(f_1, \bx) - E_\delta(f_1, \bx)\|_F^2 \right]
			&\leq
			C_3^{d-1} \mathbb{E} \left[ \|\hat{E}_\delta(f_d, \bx) - E_\delta(f_d, \bx)\|_F^2 \right]
			+
			C_4  
				\sum_{\tau = 1}^{d - 1} C_3^{\tau - 1} ,
		\end{align*}
		as $\delta \rightarrow 0$, 
		where $C_3$ and $C_4$ are constants that depend on $\dim_2(\Delta)$ and $\sigma^2$
		as well as $C_1$ and $C_2$, respectively. 
		In other words,
		the distance between the supply-chain explanation and the end-to-end explanation
		at the downstream node amplifies the upstream error by an expression that 
		can be upper bounded by an exponential of the supply chain depth $d$.
		Moreover, we show that this bound is tight. 
	\end{theorem}

The proof is given in \Cref{app:explanations_proof}.
In this result,
one can think of $\mathbb{E} [ \|\hat{E}_\delta(f_d, \bx) - E_\delta(f_d, \bx)\|_F^2 ]$ 
as the error in an end-to-end explanation and 
$\mathbb{E} [ \|\hat{E}_\delta(f_1, \bx) - E_\delta(f_1, \bx)\|_F^2 ]$
as the error in the supply-chain explanation after information is passed along a $d$-length linear supply chain.
Therefore, this result says that \textbf{the supply-chain explanation can amplify the error of 
an end-to-end explanation by a factor that is exponential in $d$, plus an additive term.} 
The root issue is that each organization does not have direct access to upstream models 
and, in the absence of such access, relies on explanations provided by upstream organizations.\footnote{Alternatively, 
downstream organizations could query upstream models many times (e.g., typically hundreds or thousands of times for LIME estimates), 
but doing so can increase API costs by multiple orders of magnitude, 
so we assume that downstream organizations would not do this unless absolutely necessary.}

Another way of viewing this result is that the empirically estimated explanations $\hat{E}_\delta(h, \bz)$ must be sufficiently close to the \emph{true} value $E_\delta(h, \bz)$ for explanations in AI supply chains to be meaningful.
This result has many implications. 
For instance, from a policy perspective, 
if model explanations are legally mandated but a model is situated in an AI supply chain, 
policymakers must ensure that either 
(1) the model's ancestor tree is not too deep (or wide), or
(2) that all upstream organizations provide explanations that are accurate within a tolerance that increases in strictness as the ancestor tree grows,
which places a strict requirement on upstream organizations.

Although our case study is centered around explanations, 
this result implies that information passing of any kind along the AI supply chain should be considered carefully.

\subsection{Empirical Evaluation}\label{sec:explanations_exp}

\begin{figure}[t!]
    \centering
    \begin{subfigure}[b]{0.32\textwidth}
        \centering
        \includegraphics[width=\textwidth]{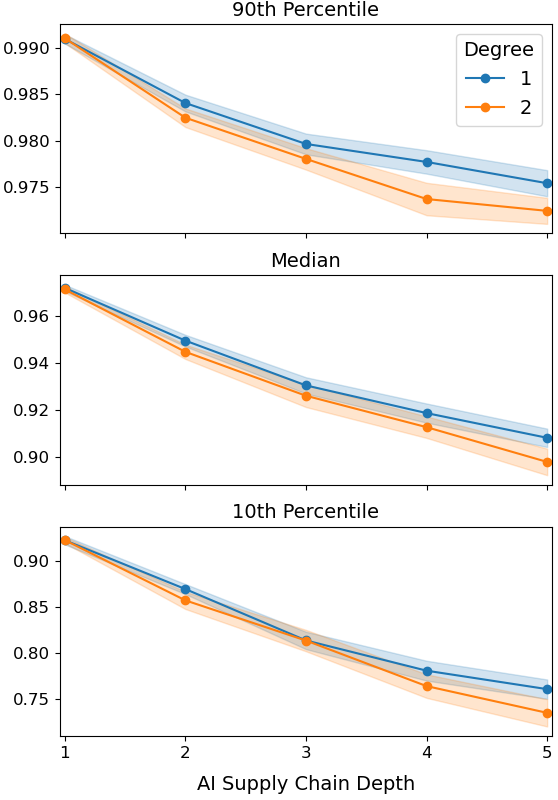}
        \vspace{-2pt}
        \caption{Cosine similarity}
        \label{fig:exp_cos_sim}
    \end{subfigure}
    \hfill
    \begin{subfigure}[b]{0.32\textwidth}
        \centering
        \includegraphics[width=\textwidth]{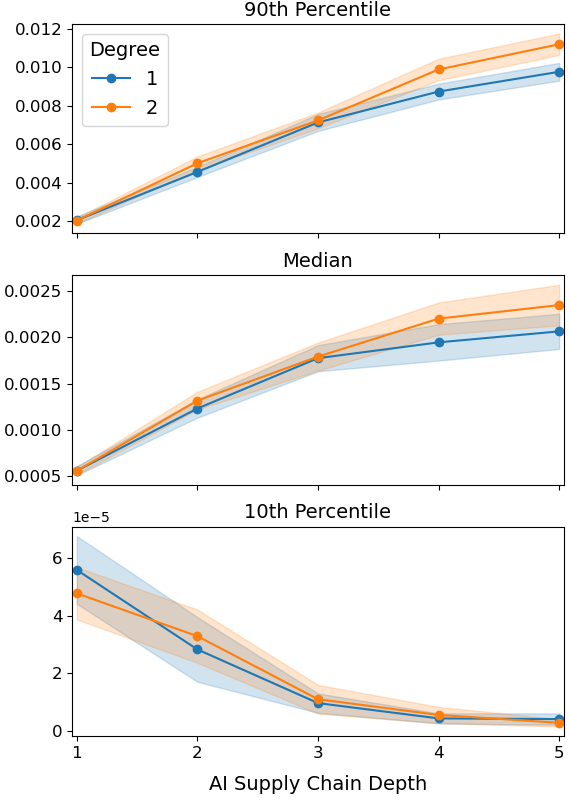}
        \vspace{-2pt}
        \caption{MSE}
        \label{fig:exp_mse}
    \end{subfigure}
    \hfill
    \begin{subfigure}[b]{0.32\textwidth}
        \centering
        \includegraphics[width=\textwidth]{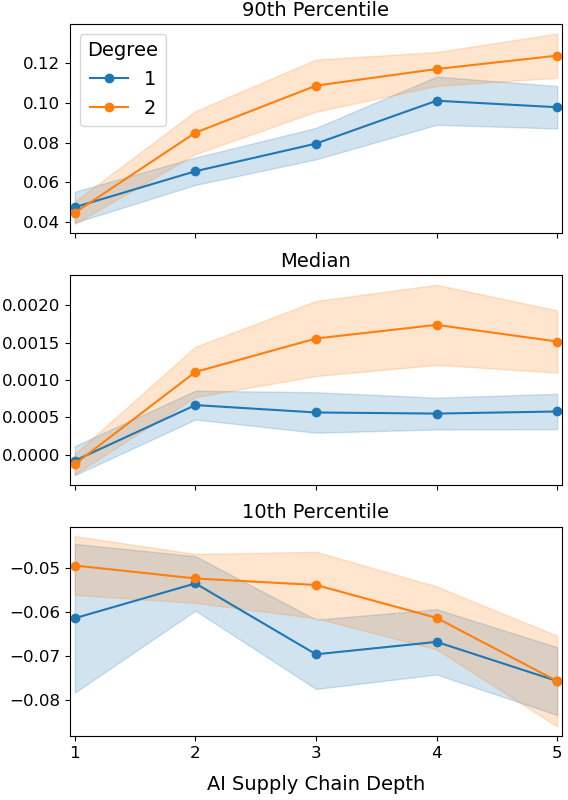}
        \vspace{-2pt}
        \caption{Recourse error}
        \label{fig:exp_rec_dist}
    \end{subfigure}
    \vspace{4pt}
    \caption{
        Simulations for \Cref{sec:explanations_exp} for 50 LIME samples and a LIME radius of 0.2.
        From left to right, 
        we plot (a) the cosine similarity between end-to-end and supply-chain explanations,
        (b) the mean-squared error (MSE) between the two explanations,
        and (c) the recourse error between the two explanations.
        In all three figures (a)-(c),
        the x-axis is the depth of the ancestor tree and the legend gives the ``width'' or degree of the ancestor tree. 
        The three subplots show the means and 95\% confidence intervals of the 
        90th percentile, 50th percentile (median), and 10th percentile of the respective metric.
    }
    \label{fig:explanations}
\end{figure}

In the previous section, we show that AI supply chains complicate the process of generating 
accurate explanations.
We now study this phenomenon experimentally, 
extending the theoretical result from a linear supply chain to one in which a node's ancestors are an inverted $m$-ary tree of depth $d$.

\paragraph{Setup.}
As before, we refer to ``end-to-end'' explanations as LIME explanations computed when given access to the entire AI supply chain,
and ``supply-chain'' explanations as those computed by combining the explanations of upstream models.
We use the chain rule to compute the supply-chain explanations (see \Cref{sec:explanation_setup}).
Our goal is to compare the end-to-end and supply-chain explanations; 
if there is information loss in the supply chain, 
the two explanations will differ. 

For the AI supply chain, 
we consider the explanations for a downstream node whose ancestor graph is an $m$-ary inverted tree of depth $d$;
doing so allows us to study different AI supply chains by varying $m$ and $d$.
At each level of the tree, 
a model passes the query of interest $\bx$ to its parents, 
then concatenates the output of its parents to $\bx$ and passes this to its own model.
Each model (i.e., node) in the ancestor graph is a simple three-layer neural network with ReLU activations and
BatchNorm after each hidden layer, 
ending with a linear layer and sigmoid activation that produces a scalar output.
The hidden dimensions double with each hidden layer. 
In order to mimic real-world model interactions,
models further upstream are larger; 
specifically, the first hidden layer of a node's model is $16$ multiplied by the ``level'' of node in the ancestor tree.
Similarly, the larger models are trained for more epochs ($15 \sqrt{\text{node level}}$) 
and on larger datasets ($1000 \sqrt{\text{node level}}$).
All models are trained with a learning rate of $0.001$ and a batch size of $32$.\footnote{Note
that the training conditions are not necessarily realistic, 
but as we are only interested in the fidelity of explanations
and not the performance of the models,
the predictive abilities of the models (even what they predict) do not affect our results.
In fact, one could run our experiment on randomly initialized models.}

We generate samples $\bx \in \mathbb{R}^{\text{featDim}}$ by sampling a mixture of two isotropic Gaussians with variance $1$, 
where the cluster centers are sampled uniformly at random within the unit hypercube.
We generate binary training labels using a second-order polynomial model with Gaussian noise, 
where coefficients are sampled uniformly in $[-1, 1]$, 
and the output is then passed through a sigmoid function and thresholded at $0.5$.

For each trial, we compare the end-to-end and supply-chain explanations for 100 different values of $\bx$. 
We run 50 trials and set featDim to 12. 
We set number of samples LIME uses to compute each explanation to 50 or 100 and the LIME radius to 0.1 or 0.2, 
with further results in \Cref{app:additional_explainability_results}.

\paragraph{Results.}
We measure three quantities of interest:
The first is the cosine distance between end-to-end and supply-chain explanations.
The second is the mean-squared error (MSE) between the two explanations (supply-chain minus end-to-end).
The third is what we refer to as the \textit{recourse error}: 
the distance one would travel along the direction indicated by the supply-chain explanation in order to change one's predictive outcome
minus the analogous quantity for the end-to-end explanation.
Intuitively, while explanations are sometimes of standalone interest, 
they often serve as mechanisms to give individuals a path to {\em recourse}. 
That is, individuals use explanations to improve their outcomes
(e.g., if an applicant is rejected, they may use the explanation to improve their qualifications and therefore future applications).  
Our definition of recourse error is inspired by the literature on counterfactual explanations, 
which examine how far one would have to move along the direction indicated by the explanation to flip one's prediction from positive to negative, or vice versa \cite{wachter2017counterfactual}.
To avoid instabilities in our analysis, we set the max value of the recourse error to 1000.
Our analysis therefore sheds light on whether explanations produced through an AI supply chain can mislead or place undue burden on individuals.

Our results are summarized in \Cref{fig:explanations}. 
For each measure, there are three subplots for the 90th, 50th, and 10th percentiles of the respective metric.
More explicitly, 
recall that we run the simulation 50 times to average over the randomness in the model training and data generation.
Moreover, recall that, for each simulation we computed explanations for 100 different values of $\bx$.
The percentiles are taken over the 100 values of $\bx$, 
and the means and 95\% confidence intervals are taken over the 50 trials.
We choose to provide these three percentiles to illustrate the ``best'', ``median'', and ``worst'' case outcomes across values of $\bx$, 
as means and medians may fail to capture the overall behavior.

Corroborating our theoretical analysis,
\textbf{the cosine similiarity between end-to-end and supply-chain explanations degrades as the supply chain length increases}, 
as can be seen by all the decreasing trends in \Cref{fig:exp_cos_sim}.
Additionally, as one may expect, \textbf{the cosine similarity also degrades with the ancestor tree's degree}. 
The MSE in \Cref{fig:exp_mse} indicates a similar trend, 
as both the 90th percentile and median values increase with depth and degree. 
Although the 10th percentile MSE seems to improve, note that the scale of this plot's y-axis is very small.
Finally, 
we also examine the recourse error in \Cref{fig:exp_rec_dist}.
It shows that the error increases with the depth and degree of the supply chain, 
which extends beyond our theoretical findings
(note that the 90th percentile recourse error scale is larger than the other two percentiles).
Moreover, if we study the sign of the recourse error (which is the supply-chain recourse distance minus the end-to-end recourse distance),
we find that \textbf{the recourse distance of the supply-chain explanation is consistently larger than that of the end-to-end explanation}, 
across all three percentiles. 

Taken with the cosine similarity result, 
\Cref{fig:exp_cos_sim} and \Cref{fig:exp_rec_dist} indicate that \textbf{if a decision subject seeks to change their outcome under an AI-driven decision,
the supply-chain explanation points them in a different direction than an end-to-end explanation would, 
and this causes decision subjects to have to exert more effort than they would under the end-to-end explanation.}

		\section{Case Study 2: Upstream Choices Affect Downstream Outcomes}
\label{sec:fairness}

In the previous section, we examine how communication along the AI supply chain can lead to misleading information downstream. 
In this section, we turn our attention to how decisions made upstream  can fundamentally restrict downstream actors.
We study a setting in which an upstream model is designed to meet some notion of algorithmic fairness. 
We then show that \textbf{fine-tuning on this model restricts downstream models in their ability to achieve their own performance-fairness objectives}. 
Our result goes beyond showing that downstream models inherit the upstream notion of fairness. 
In fact, we show \textbf{the downstream model can \emph{undo} upstream fairness, but at the cost of other desiderata}.
Thus, training upstream models to satisfy an attribute does not guarantee that downstream models
will inherit that attribute but can restrict the downstream developer in other unexpected ways.
Although we study algorithmic fairness, this result does not indicate that implementing fairness is restrictive;
rather, we use fairness as a concrete example but believe many design choices (even the lack of an explicit choice as a choice in and of itself) have downstream effects.\footnote{
    Although we focus on fairness, we believe our findings signal more broadly that upstream decisions have unexpected downstream repercussions. Notably, our results do not show that upstream decisions are simply \emph{reflected} in downstream models (i.e., upstream fairness appears downstream); rather, upstream decisions can be ``undone'' but at the cost of other considerations.}

\subsection{Problem Setup}
Consider an upstream base model $f_p$ and a downstream model $f_v$, 
where $f_v$ is fine-tuned on $f_p$. 
Suppose both $f_v$ and $f_p$ take in inputs $\mathbf{x} \in \bbR^\rho$, 
which are realizations of the random variable $X$, and output $y \in \bbR$.\footnote{We adopt the convention that capital letters denote random variables and lowercase letters denote realizations.}
For instance, 
$\bx$ could be a vector of features representing an individual that the downstream model seeks to evaluate, 
and instead of using the off-the-shelf base model $f_p$, 
the downstream actors fine tunes it on specialized data to produce $f_v$.

\paragraph{Upstream design choices.}
Model developers inevitably make design choices when training their models. 
These choices are not necessarily optimal for downstream actors, 
as developers cannot tailor their base models to suit all possible downstream use cases
(unless they are providing custom models).  
In our analysis, 
we use the abstraction that the upstream design choice can be represented as a \emph{conditional independence desideratum}. 
That is, 
we suppose $f_p$ is trained such that $f_p(X)$ is (approximately) conditionally independent of $X_1$ given random variable $Z \in \mathcal{Z}$, 
i.e., 
$$f_p(X) \perp X_1 \, | \, Z,$$
where $X = (X_1, X_2, \hdots)$, 
and the randomness is given by the training data distribution
For example, $Z$ could be $(X_2, \hdots, X_d)$ or $Z$ could be the empty set.

We adopt this setup because conditional independence encompasses a family of requirements that appear not only in algorithmic fairness but other types of ``structured'' settings, 
such as causal inference on structural causal models. 
It is thus a natural class of requirements that an upstream developer may consider.
For instance, statistical parity and equalized false positive rates are both notions of algorithmic fairness that can be expressed as conditional independence requirements \cite{barocas2017fairness}. 
Statistical parity for sensitive attribute $X_1$ is equivalent to setting $Z = \emptyset$, and equalizing odds across sensitive groups determined by $X_1$ is equivalent to setting $Z = Y$, where $Y$ is the true outcome (e.g., how well a job applicant would perform if they were hired).
Intuitively, $X_1 \perp f_p(X) \, \mid \, Z$ asks that, given information $Z$, knowing the value of a sensitive attribute (e.g., race or sex) $X_1$ provides no additional information about $f_p (X)$.

We examine the implications of fine-tuning $f_v$ on $f_p$ when $f_p$ satisfies this conditional independence. 
In the next two sections, 
we provide a theoretical characterization, 
followed by a more general empirical evaluation.

\subsection{Conditional independence affects embeddings}

We now show that upstream design choices affect downstream models 
in ways that are not immediately obvious.
Although one might expect that imposing a constraint on upstream model 
$f_p$ would lead the downstream model $f_v$ to inherit that same constraint,
we show (as is consistent with the empirical findings in the literature 
\cite{salman2022does,steed2022upstream,yuan2024closer,hu2025unlearning})
that one can actually ``undo'' the upstream constraint, 
but it has implications on other objectives that the downstream developer has.

In this section, we begin with a theoretical result. 
To study this setting analytically, 
we model $f_p$ as
a linear combination of basis functions:
\begin{align}
    f_p(\mathbf{x}) = \sum_{i=1}^N {\phi}_i(\mathbf{x}) {w}_i, \label{eq:basis_function_f}
\end{align}
where $\mathbf{\mathbf{x}} = ({x}_1, {x}_2, \ldots, {x}_\rho) \in \mathbb{R}^\rho$, 
${\phi}_i : \mathbb{R}^\rho \to \mathbb{R}$, 
${w}_i \in \mathbb{R}$,
$\mathbf{w} = (w_1, \ldots, w_N) \in \mathbb{R}^N$,
$\boldsymbol{\Phi} = (\phi_1, \ldots, \phi_N) \in \mathbb{R}^{N \times \rho}$.
We assume that $\phi_i$'s (and thus $f_p$) are deterministic.
Similar models appear in kernel methods \cite{scholkopf2002learning}, function approximation \cite{murphy2012machine, hornik1989multilayer, cybenko1989approximation}, and deep learning theory \cite{bengio2017deep}. 
Such models are popular because it is a universal function approximator as $N \rightarrow \infty$ under sufficiently rich basis functions. 
Intuitively, one can think of the basis functions as embeddings and $N$ as a proxy for model size.

We model fine-tuning a downstream model $f_v$ on an upstream model $f_p$ 
as learning a linear function on the embeddings of the upstream model $f_p$. 
As the basis functions $\boldsymbol{\phi}_i$ represent the embeddings of $f_p$, 
the fine-tuned model is given by $f_v(\mathbf{x}) = \sum_{i=1}^N \boldsymbol{\phi}_i(\mathbf{x}) {r}_i,$
where the coefficients $r_i \in \bbR$ are learned from fine-tuning.

\begin{theorem}\label{thm:fairness}
    Suppose the base model is trained to achieve approximate independence of $f_p(X)$ and $X_1$ given $Z$
        such that
        there exists $\varepsilon_p > 0$ for which
        \begin{align*}
                \left| \frac{\partial}{\partial x_1} \mathbb{E}_{D_p}[f_p(X) \mid X_1 = x_1, Z = z] \right| \leq \varepsilon_p, 
        \end{align*}
        for all $x_1 \in \mathbb{R}$ and $z \in \mathcal{Z}$.
        Let $\varepsilon_v$ denote the analogous value achieved by $f_v$ 
        (i.e., how close $f_v(X)$ is to conditional independence from $X_1$ given $Z$).
        Note that this is an appropriate condition for approximate conditional independence, 
        since $f_p$ is a binary classifier.\footnote{For regression, one would require instead that there exists an $\varepsilon' > 0$ such that 
        $\left| \frac{\partial}{\partial x_1} {P}[f_p(X) = y \mid X_1 = x_1, Z = z] \right| \leq \varepsilon_p$
        for all $x_1 \in \mathbb{R}$, $z \in \mathcal{Z}$, and $y \in \mathbb{R}$.}
    Similarly, suppose the downstream model is trained to achieve approximate independence of $f_v(X)$ and $X_2$ given $Z$ such that
    \begin{align*}
            \left| \frac{\partial}{\partial x_2} \mathbb{E}_{D_v}[f_p(X) \mid X_2 = x_2, Z = z] \right| \leq \eta_v, 
    \end{align*}
    for all $x_2 \in \mathbb{R}$ and $z \in \mathcal{Z}$.
    Then,
    as long as $\varepsilon_p, \eta_v > 0$ and the bounds above are tight for some value of $(x_1, z)$ and $(x_2, z)$,
    then there exist functions $g$, $S$, and $S'$ such that 
    such that
    \begin{align}
        \frac{\varepsilon_v}{
            \varepsilon_p} 
        \cdot
        C_5(\mathbf{w}, \boldsymbol{\Phi})
        \leq
        \| \mathbf{r} \| 
        &\leq 
        \eta_v
        \cdot
        C_6(\mathbf{r}, \boldsymbol{\Phi}) \label{eq:fairness_bounds}
        ,
    \end{align}
    for constants $C_5$ and $C_6$ that depend on $\mathbf{w}$ and $\mathbf{r}$, respectively,
    and both constants can depend on $\boldsymbol{\Phi}$ and the types of upstream/downstream constraints. 
\end{theorem}
A proof is given in \Cref{sec:proof_fairness}.
This result shows that a design choice made on the upstream model $f_p$
(as captured by $\varepsilon_p$) affects the downstream, fine-tuned model $f_v$
(as captured by $\| \mathbf{r} \|$, $\varepsilon_v$, and $\eta_v$). 
Concretely, 
we can make three observations from the result (and corresponding proof). 
First, 
the downstream model $f_v$ does not necessarily inherit the upstream conditional independence. 
In fact, $f_v$ can ``undo'' it, which would be captured by $\varepsilon_v$ being large compared to $\varepsilon_p$.
This leads to our second observation,
which is that the upstream conditional independence (with respect to $X_1$) affects 
$f_v$'s ability to achieve its conditional independence with respect to $X_2$.
For example, 
even if $f_v$ may wish to ``undo'' the upstream conditional independence with respect to $X_1$, 
doing so implies increasing $\varepsilon_v$, which raises the lower bound on $\| \mathbf{r} \|$.
However, the right-hand side of \Cref{eq:fairness_bounds} shows that $\| \mathbf{r} \|$ is also 
bounded above by $\eta_v$ multiplied by a constant that (as shown in the proof) does not
depend on $\| \mathbf{r} \|$, 
but this may not be compatible with $f_v$'s desire to achieve their own
downstream design desideratum by keeping $\eta_v$ small. 
(Although one may be able to satisfy both conditional independences by letting
$\eta_v$ and $\varepsilon_v$ \emph{both} be small, 
this is often impossible in practice because $\eta_v$ and $\varepsilon_v$ are often
inversely related. For example, it is well known that various notions of algorithmic fairness are 
impossible to achieve simultaneously \cite{chould2017fair}).
Third, the downstream model is restricted in its choice of model parameters $\mathbf{r}$
if it wishes to fine-tune \emph{and} satisfy its own desideratum $\eta_v$. 
Because the downstream developer cannot change the upstream constraint $\varepsilon$, 
it \emph{must trade off between what it can control: its own desideratum and $f_v$'s performance}.

In other words, 
this result shows that 
\textbf{(i) the downstream model does \emph{not} necessarily inherit the 
    upstream constraint and can in fact reverse it, 
    but (ii) even so, the upstream constraint restricts the downstream model
    and imposes a trade-off between the downstream's performance and own desideratum}
    (or, more accurately, imposes a trade-off that is stronger than what would naturally exist).
Fact (i) is consistent with the literature showing that
upstream unlearning or fairness can be undone with fine-tuning
that is often interpreted to mean that upstream constraints do not
have downstream impact. 
However, fact (ii) implies that, if we analyze multiple properties
of $f_v$ simultaneously instead of separately, we will observe the ``footprint''
of the upstream constraint. 

In the next section, 
we perform this analysis empirically by studying the Pareto frontiers of 
downstream models that are fine-tuned on different upstream models
and with different downstream desiderata.

\subsection{Empirical Evaluation}\label{sec:empirical_fairness}

\begin{figure}[t!]
        \centering
        \includegraphics[width=\textwidth]{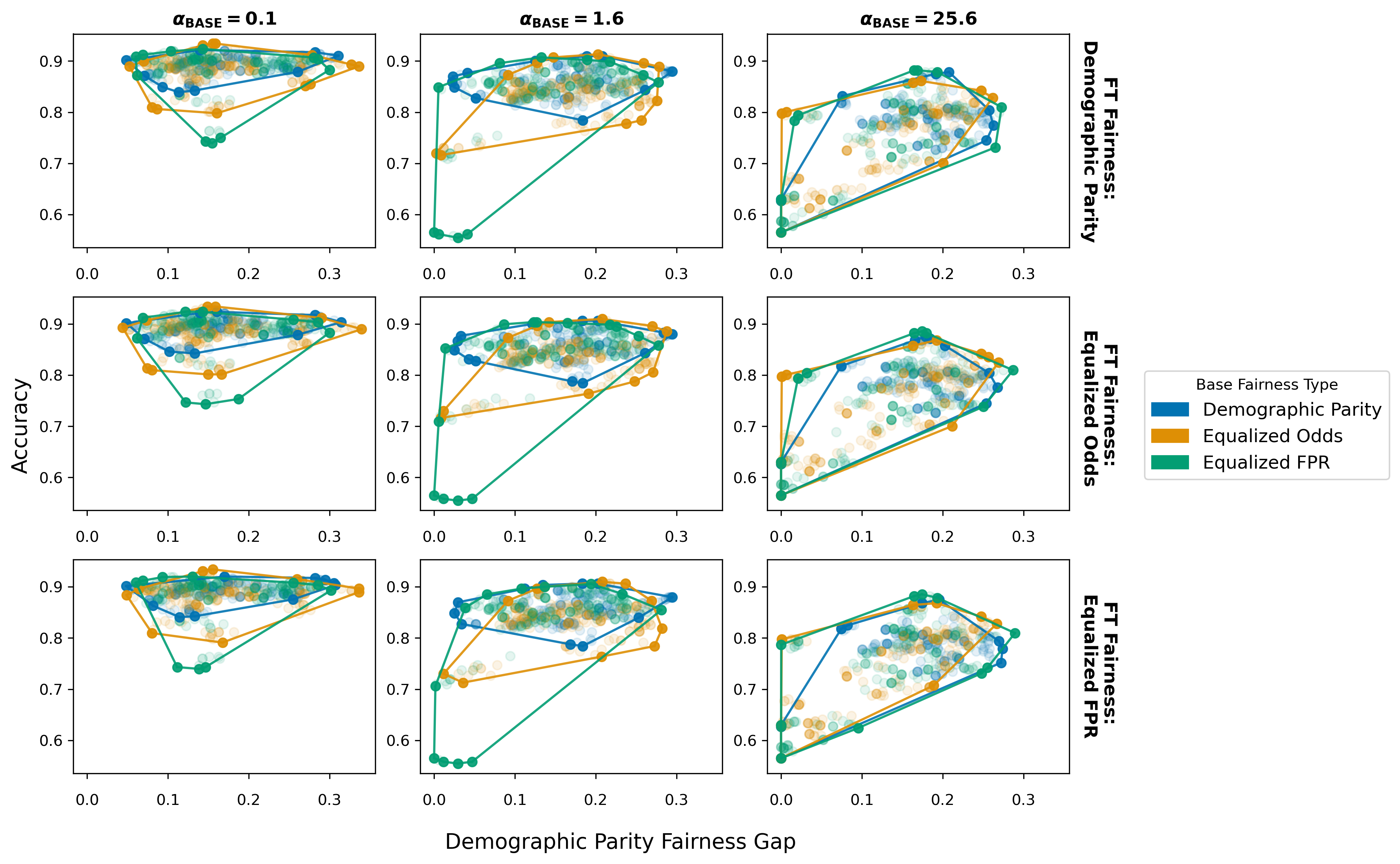}
    \caption{
        Results for Section \ref{sec:empirical_fairness}.
        Every point corresponds to a downstream (fine-tuned) model. 
        On the x-axis, we plot the model's demographic parity fairness gap
        (the absolute difference between the positive classification rates of the two sensitive groups),
        and on the y-axis, we plot the model's accuracy.
        Each column corresponds to a different base model regularization constant
        (i.e., the strength of the fairness regularization used to train the base model).
        Each row corresponds to a different type of downstream fairness criterion.
        The color of each point corresponds to the type of upstream fairness regularizer (used to train the base model).
        We show the convex hull of the points (that correspond to the same color, or base regularizer)
        where the top-left portion of the convex hull corresponds to the empirical performance-fairness Pareto frontier.
        To sweep out the Pareto frontier and elicit downstream models with different performance-fairness characteristics, 
        we vary the fine-tuning regularization constant $\alpha_v$.
}
    \label{fig:fairness_dp}
\end{figure}

\begin{figure}[t!]
        \centering
        \includegraphics[width=\textwidth]{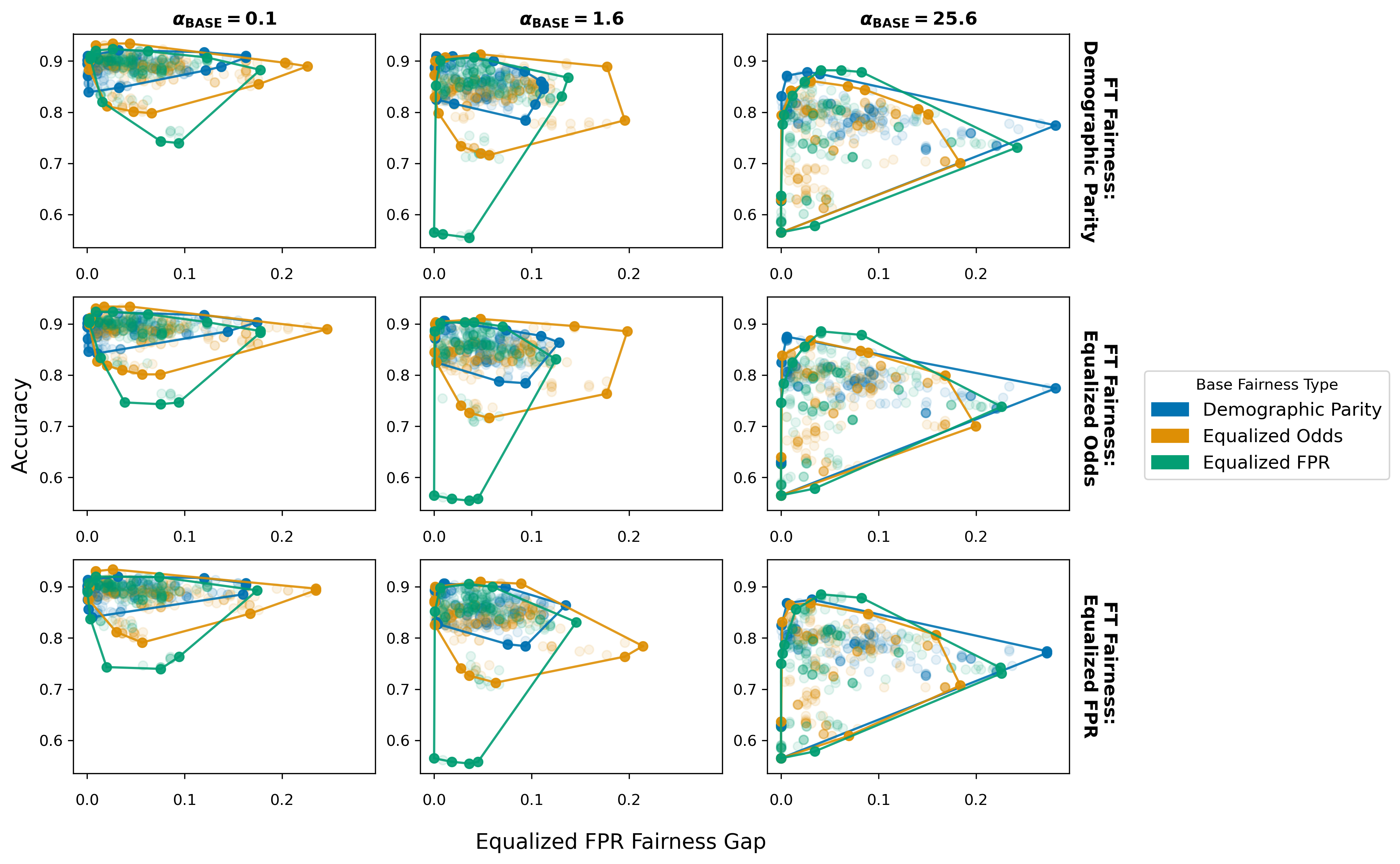}
    \caption{
        Results for Section \ref{sec:empirical_fairness}.
        In each plot, 
        every point corresponds to a downstream (fine-tuned) model. 
        On the x-axis, we plot the model's equalized FPR fairness gap
        (the absolute difference between the FPR of the two sensitive groups),
        and on the y-axis, we plot the model's accuracy.
        The remanining figure layout is identical to that in Figure \ref{fig:fairness_dp}.
    }
    \label{fig:fairness_fpr}
\end{figure}

In the previous section, we show that design choices made for base models propagate
to downstream models, and that the effect is not straightforward. 
Downstream models do not simply inherit the same properties; rather, our result 
indicates that one can surface the upstream ``footprint''
by examining the trade-offs that downstream models must make between their own design objectives. 
In this section, we instantiate constraints as fairness criteria
and study the effect of upstream design choices on downstream models,
using performance-fairness Pareto frontiers to examine the trade-offs made downstream.

\paragraph{Setup.}
Mirroring our theoretical analysis,
we have a base model $f_p$ trained on dataset $D_p$ 
and a downstream model $f_v$ fine-tuned on $f_p$ using dataset $D_v$.
Both $f_p$ and $f_v$ are binary classifiers, 
and they are trained to minimize the binary cross-entropy loss 
plus a fairness regularization term, 
moderated by constant $\alpha$,
such that the base model seeks to minimize
\begin{align*}
    \mathcal{L}_{\text{BASE}}(f_p, D_p) = \text{BCE}(f_p, D_p)
         + \alpha_p \cdot \mathcal{R}_{\text{fairness}}(f_p, D_p) \text{,}
\end{align*}
and the downstream model seeks to minimize
\begin{align*}
    \mathcal{L}_{\text{FT}}(f_v, D_v) = \text{BCE}(f_v, D_v)
         + \alpha_v \cdot \mathcal{R}_{\text{fairness}}(f_v, D_v) \text{.}
\end{align*}
Both the base and fine-tuned models use ResNet18 as their architectures
and are trained and tested on images
from the Waterbirds dataset \cite{koh2021wilds}.
Waterbirds is a dataset of images that contain either ``land'' or ``water'' birds
on ``land'' or ``water'' backgrounds. 
Thus, previous works \cite{koh2021wilds,liu2021just,sagawa2019distributionally,shah2023modeldiff} 
have used Waterbirds to study algorithmic fairness
since the bird label should not be sensitive to the background label,
but the labels are correlated (as water birds are more likely to be on water backgrounds, 
and vice versa).
In our experiment, both the base and fine-tuned models are trained 
to predict the bird label from the image, 
with the background as the sensitive attribute.\footnote{\label{footnote:balancing}
    Importantly, the dataset is not balanced. 
    We resample to balance the backgrounds and classes  as follows: 
    40\% water birds on water background; 
    20\% water birds on land; 
    20\% land birds on water; 
    20\% land birds on land. 
    To avoid overlap between the base training/testing and fine-tuning training/test sets,
    we further re-shuffle and split the dataset such that 
    the base-train, base-test, fine-tuning-train, and fine-tune-test sets
    constitute approximately 60\%, 10\%, 20\%, and 10\% of the eligible data, respectively. 
    Note that although we could have allowed for distribution shift (between train and test sets, 
    or even between the base and fine-tuning sets),
    we intentionally chose not to do so to isolate the effect of the upstream model
    on the downstream model and remove possible confounding factors, like distribution shift. 
}

To analyze fairness, 
we use three different notions of algorithmic fairness:
demographic parity, 
equalized false positive rate (FPR), and equalized odds.
Their formal definitions can be found in \citep{barocas2017fairness};
they correspond to asking that 
the positive and negative classification rates for groups differing only by sensitive attribute are the same;
the false positive rates across sensitive groups are the same;
and finally that both the false and true positive rates are the same;
respectively.\footnote{
    To avoid discontinuities in the loss function while training,
    we use a smoothed version of the fairness regularizers
    (i.e., avoid indicators of whether $f(x) = y$ that appear in 
    expressions of demographic parity, equalized FPR, and equalized odds).
}

We train the base model (which is initialized using weights pre-trained on ImageNet)
for 40 epochs, with a batch size of 256 and learning rate of 0.001.\footnote{
    We use a ResNet18 model has already been pre-trained on ImageNet
    Although our base model itself is a fine-tuned model, 
    our findings still hold because the final model is still trained 
    to satisfy the desired criteria. 
    In fact, this approach matches the setting of supply chains well;
    in practice, a downstream model can become a base model (upstream model) 
    for another model further downstream.}
For fine-tuning, we use the same hyperparameters, 
except that fine-tuning takes place only over the final layer of the model.
We repeat base model training 20 times under different random seeds. 
For each random seed, we train the base model under different values of $\alpha_{p}$ ranging from 0 to 25.6. 
We then fine-tune each base model on $D_v$ 
for values of $\alpha_{p}$ ranging from 0 to 25.6. 
In total, we trained over 10,000 models.

\paragraph{Results.}
Figures \ref{fig:fairness_dp} and \ref{fig:fairness_fpr} 
summarize our findings.
In each, 
we plot the accuracy on the y-axis and the fairness gap (larger values denote greater unfairness) on the x-axis, 
where the fairness notion in Figure \ref{fig:fairness_dp} is demographic parity
and in Figure \ref{fig:fairness_fpr} is equalized FPR.
Each point in the plot corresponds to a fine-tuned model $f_v$.
To obtain a range of fine-tuned models with different performance-fairness characteristics, 
we vary the fine-tuning regularization constant $\alpha_v$.
The fairness regularizer used to train the base model is denoted by each point's color.
Each column corresponds to a different base model regularization constant
(i.e., the strength of the fairness regularizer used to train the base model).
Each row corresponds to a different type of downstream (fine-tuning) 
fairness criteria/regularizer. 

Put simply, 
each plot shows the range of achievable downstream models
under different upstream constraints (i.e., the base model's fairness notion).
The columns correspond to the strength of the upstream constraint (higher values indicate stronger constraints). 
The rows correspond to the downstream desideratum (which may not match the upstream one).
Although we choose to visualize all points (i.e., all fine-tuned models), 
the top-left curve of each plot corresponds to the empirical performance-fairness Pareto frontier.

There are three key takeaways from both figures.
First, \textbf{the stronger the upstream constraint, 
the greater the performance-fairness trade-off on the downstream model},
as can be observed by pareto frontier moving down and to the right
as we move from left to right within each row.
Second, within each column, 
there is not much variation across rows compared variation across colors within each plot, 
indicating that \textbf{the downstream constraint (row) has a limited effect on the Pareto frontier 
compared to the upstream constraint (color)}. 
Third, there is some (though limited) variation across rows, 
indicating that the downstream constraint does have some effect on the 
achievable performance and fairness of the downstream model.
These trends are consistent across both figures. 
Between figures, accuracy and equalized FPR are more compatible
than accuracy and demographic parity,
which is consistent with the figures showing a greater trade-off for the latter. 
We do not see any consistent trends when examining the Pareto frontiers 
for matching upstream and downstream 
constraints versus differeing upstream and downstream constraints, 
which is not necessarily expected, 
but consistent with our theoretical finding since the constants $C_5$ and $C_6$
depend on the types of constraints.

		\section{Conclusion}\label{sec:conclusion}
In this paper, we call attention to the emergence of AI supply chains and offer a model through which to interrogate their implications on learning. 
Through our case studies, we show that developing machine learning systems across AI supply chains 
raises challenges that have technical and legal ramifications.
Our findings underscore the need for new frameworks and methodologies that account for the emergence of AI supply chains,
including improving evaluation metrics, auditing mechanisms, and regulatory oversight.

	\section*{Acknowledgments}
	
	We would like to thank Percy Liang, Daniel E. Ho, Helen Nissenbaum, Hongseok Namkoong, 
	Frederike Zufall, Jat Singh, Max Simchowitz, Susan Silbey,
	and others for fruitful conversations on AI supply chains. 
	The authors gratefully acknowledge funding from the Open Philanthropy Project, 
	National Science Foundation (NSF) grant CNS-1955997
	and the Air Force Research Laboratory (AFOSR) grant FA9550-23-1-0301.

	\printbibliography

	\clearpage

	\appendix
	\section*{\LARGE Appendix}

	\section{Proof of Section \ref{prop:explanation}}\label{app:explanations_proof}

	Before we provide the proof, consider the following example, 
	which provides some intuition for the result.

	\begin{example}
		Consider a linear supply chain of depth $d$,
		composed of functions $h_i$.
		Each node in the supply chain is able to compute an (approximate) explanation 
		$\hat{E}_\delta(h_i, \bz_i)$ that explains how its own model $h_i$ behaves in 
		terms of its inputs $\bz_i = (\bx, \by_{i+1})$, where $\by_{i+1} = f_{i+1}(\bx)$ is the output of $i$'s parent.		
		Given a feature vector $\mathbf{x}$ (e.g., applicant's features), 
		each node $i$ computes 
		its output $f_i(\mathbf{x}) = h_i(\bx, y_{i+1} = f_{i+1}(\mathbf{x}))$,
		where $f_{i+1}$ is the function computed by its direct ancestor, or parent. 
		Analogously, node $i$ computes the explanation $\widehat{E}_\delta(f_i, \mathbf{x})$ relative to $\bx$ 
		by taking the explanation $\hat{E}_\delta(f_{i+1}, \mathbf{x})$ provided by its parent, 
		and returning $\hat{E}_\delta(h_i, \bz_i)^\top E_\delta(f_{i+1}, \mathbf{x})$,
		which is the result of applying the ``chain rule'' from \Cref{sec:explanation_setup} to a linear supply chain.
	\end{example}

	Suppose, for the purpose of illustration,
	that in the context of the example above, 
	we have $h_i(\mathbf{z}) = \mathbf{z}$ for all the nodes in the supply chain.
	Note that this function is, in fact, linear,
	making locally linear explanations ``faithful'' to the underlying model;
	we will show that, even in this case, error propagation along the supply chain
	can lead to mis-estimates of the true end-to-end explanation.

	Recall that locally linear approximations cannot typically be computed exactly 
	and are thus approximated empirically by sampling the $\delta$-neighborhood around $\bz$.
	Thus, suppose $\hat{E}_\delta(h_i, \mathbf{z}) = E_\delta(h_i, \bz) + \Delta_i$,
	where $\Delta_i$ is a diagonal matrix whose entries are drawn i.i.d. from an 
	arbitrary non-degenerate distribution $D$ 
	with values bounded between $[-1, 1]$ for all $i$, 
	(where non-degenerate means that the variance of the distribution is non-zero). 
	Since $h_i(\bz) = \bz$ is linear, 
	$\widehat{E}_\delta(h_i, \bz) := \prod_{i=1}^d (\mathbf{I} + \Delta_i)$
	for all $\bz$.

	This explanation will become increasingly far from the true 
	explanation as $d$ increases. 
	As such, the distance between the ``supply-chain'' explanation and the ``end-to-end'' one 
	(which can be interpreted as an explanation for a single-node supply chain)
	will increase with $d$. 
	In particular, the eigenvalues of 
	$\widehat{E}_\delta(h_i, \bz_i)$ satisfy, for any $k, \ell \in [\text{dim}(\bz)]$ and $k \neq \ell$,
	\begin{align*}
		\mathbb{E}\left[\log\left(\frac{\lambda_k}{\lambda_\ell}\right)^2\right] 
		&\geq 
		\text{Var}\left[\log\left(\frac{\lambda_k}{\lambda_\ell}\right)\right] \\
		&= 
		\text{Var}\left[\log\left(\frac{\prod_{i=1}^d (1 + \Delta_i[k,k])}{\prod_{i=1}^d (1 + \Delta_i[\ell,\ell]) }\right)\right] \\
		&= d \cdot V,
	\end{align*}
	where $V = \text{Var}_{T, T' \sim D}[\log(\frac{1+T}{1+T'})] > 0$.
	Thus, there exist $k$ and $\ell$ such that $\lambda_k / \lambda_\ell$
	grows approximately $\exp(d V / 2)$. 
	Most importantly, this occurs even though $E_\delta(h_i, \bz) = \mathbb{I}$'s eigenvalues are all equal, 
	and therefore the ratio between all eigenvalues does not grow with $d$.
	In other words, even if the true explanation says that $h_i$ values all dimensions (or directions)
	of its inputs equally, the supply-chain explanation will compound the ``error'' in 
	$\hat{E}$'s in a way that scales with the supply chain depth $d$.
	Below, we prove a similar statement that holds more generally. 

	We now provide the full proof of \Cref{prop:explanation}

	\begin{proof}
		Let $i = 1, 2, \ldots, d$ be the index of the node in the (linear) supply chain,
		where $i = d$ is the most upstream node.
		Recall that, from the chain rule for explanations that actors in our setup use (\Cref{sec:explanation_setup}),
		the explanation at node $i$ relative to $\bx$ is given by
		\begin{align*}
			\hat{E}_\delta(f_i, \bx) 
			&= 
			\sum_{j \in \text{pa}(i)} 
			\hat{E}_\delta(h_i, \bz_i)^\top 
			\hat{E}_\delta(f_j, \bx) .
		\end{align*}
		Thus, for a linear supply chain where there is only one parent $i+1$ for node $i$, 
		\begin{align*}
			\hat{E}_\delta(f_i, \bx) 
			&= 
			\hat{E}_\delta(h_i, \bz_i)^\top 
			\hat{E}_\delta(f_{i+1}, \bx) ,
		\end{align*}
		where $i+1$ is $i$'s parent.

		\paragraph{Error.}
		We begin by characterizing the ``error'' of the supply-chain explanation.
		Using the above and by the definition of $\Delta_i$ given in the theorem statement,
		the supply-chain explanation at node $i$ is given by
		\begin{align*}
			\hat{E}_\delta(f_i, \bx) 
			&= 
			\left( 
				{E}_\delta(h_i, \bz_i) + \Delta_i
			\right)^\top 
			\hat{E}_\delta(f_{i+1}, \bx) .
		\end{align*}
		Subtracting the end-to-end explanation ${E}_\delta(f_i, \bx)$ from both sides, we have
		\begin{align}
			\hat{E}_\delta(f_i, \bx) - {E}_\delta(f_i, \bx)
			&= 
			\left( 
				{E}_\delta(h_i, \bz_i) + \Delta_i
			\right)^\top 
			\hat{E}_\delta(f_{i+1}, \bx)
			- {E}_\delta(f_i, \bx) \nonumber
			\\
			&= 
			{E}_\delta(h_i, \bz_i)^\top 
				( \hat{E}_\delta(f_{i+1}, \bx) - {E}_\delta(f_{i+1}, \bx)) \nonumber
			\\
			& \hspace{20pt} +
			( {E}_\delta(h_i, \bz_i)^\top {E}_\delta(f_{i+1}, \bx)
			- 
			{E}_\delta(f_i, \bx) ) \nonumber
			\\
			& \hspace{20pt} + 
			\Delta_i^\top \hat{E}_\delta(f_{i+1}, \bx) . \label{eq:exp_proof1}
		\end{align}
		Using the Taylor series approximation,
		 ${E}_\delta( g, \mathbf{w} ) = J_g ( \mathbf{w} ) + O(\delta)$ where $J_g(\mathbf{w})$ is the Jacobian 
		of $g$ at $\mathbf{w}$ as long as $g$ is locally smooth around $\bw$ and differentiable at $\bw$,
		and we use $O(\delta)$ to denote that all entries grow (at most) linearly with $\delta$.
		Moreover, for any differentiable $g$ and $g'$ where $f(\bx) = g'(g(\bx))$,
		the Jacobian \(J_f(x)\) satisfies
		$
		J_f(x)
		=
		J_{g'}\left(g(x)\right)
		J_g(x)
		$.
		Therefore, the second term in \Cref{eq:exp_proof1} can be simplied such that
		\begin{align*}
			\hat{E}_\delta(f_i, \bx) - {E}_\delta(f_i, \bx)
			&= 
			{E}_\delta(h_i, \bz_i)^\top 
				( \hat{E}_\delta(f_{i+1}, \bx) - {E}_\delta(f_{i+1}, \bx))
		    + 
			\Delta_i^\top \hat{E}_\delta(f_{i+1}, \bx) 
			+ 
			O(\delta) .
		\end{align*}

	\paragraph{Upper bound.}
	Let	$Z_i := \hat{E}_\delta(f_i,\mathbf{x}) - E_\delta(f_i,\mathbf{x})$.
	Rewriting \Cref{eq:exp_proof1},
	\begin{align*}
	Z_i 
	&= E_\delta(h_i, \mathbf{z}_i)^\top \Bigl(\hat{E}_\delta(f_{i+1}, \mathbf{x}) - E_\delta(f_{i+1}, \mathbf{x})\Bigr)
	+ \Delta_i^\top  \hat{E}_\delta(f_{i+1}, \mathbf{x})
	+ O(\delta) \\
	&= \underbrace{ E_\delta(h_i, \mathbf{z}_i)^\top Z_{i+1} }_{A_i}
	+ \underbrace{ \Delta_i^\top  \hat{E}_\delta(f_{i+1}, \mathbf{x}) }_{B_i}
	+ O(\delta) ,
	\end{align*}
	we have
	\begin{align*}
		\|Z_i\|_F^2 
		&=
		\left\| A_i + B_i + O(\delta)\right\|_{F}^2.
		\\
		&=
		\|A_i\|_F^2 
		+
		\|B_i\|_F^2 
		+
		\|O(\delta)\|_F^2
		+
		2 \langle A_i, B_i\rangle_F
		+
		2 \langle A_i, O(\delta)\rangle_F
		+
		2 \langle B_i, O(\delta)\rangle_F.
	\end{align*}
	Next, we take the expectation of both sides and simplify. 
	To simplify, we make four observations.
	(1) Since $A_i = E_\delta(h_i,\mathbf{z}_i)^\top Z_{i+1}$ and $E_\delta(h_i,\mathbf{z}_i)$ is deterministic,
	$$
	\mathbb{E}\left[\|A_i\|_F^2\right]
	\leq
	\| E_\delta(h_i,\mathbf{z}_i)\|_2^2
	\cdot
	 \mathbb{E} \left[ \|Z_{i+1}\|_F^2\right].
	$$
	(2) Since $B_i = \Delta_i^\top  \hat{E}_\delta(f_{i+1},\mathbf{x})$ 
	(and $\Delta_i$ has i.i.d.\ mean‐zero entries with variance $\sigma^2$ that are independent of any upstream quantities),
	\begin{align*}
		\mathbb{E}\left[\|B_i\|_F^2\right]
		&= 
		\mathbb{E} \left[\| \Delta_i^\top \hat{E}_\delta(f_{i+1},x) \|_F^2 \right]
		\\
		&= 
		\mathbb{E}
		\left[\|
			\Delta_i^\top E_\delta(f_{i+1},x)
			+
			\Delta_i^\top Z_{i+1}
			\|_F^2
		\right]
		\\
		&= 
		\mathbb{E}
		\left[\|
			\Delta_i^\top E_\delta(f_{i+1},x)
		\|_F^2
		\right]
		+
		\mathbb{E}
		\left[\|
			\Delta_i^\top Z_{i+1}
			\|_F^2
		\right]
		+
		2 \mathbb{E}\left[
			\langle \Delta_i^\top E_\delta(f_{i+1},x) , \Delta_i^\top Z_{i+1} \rangle_F
		\right]
		\\
		&= 
		\text{dim}_2(\Delta_i) \sigma^2 \left(
		\|
			E_\delta(f_{i+1},x)
		\|_F^2
		+
		\mathbb{E}
		\left[\|
			Z_{i+1}
			\|_F^2
		\right]
		\right)
		+
		2 \mathbb{E}\left[
			\langle \Delta_i^\top E_\delta(f_{i+1},x) , \Delta_i^\top Z_{i+1} \rangle_F
		\right]
		\\
		&= 
		\text{dim}_2(\Delta_i) \sigma^2  
		\left(
			\|
				E_\delta(f_{i+1},x)
			\|_F^2
		+ 
			\mathbb{E}
			\left[\|
				Z_{i+1}
				\|_F^2
			\right]
		\right) ,
	\end{align*}
	where the fourth equality follows from the independence of $\Delta_i$ and all upstream quantities, 
	that $\Delta_i$ has i.i.d. mean-zero entries with variance $\sigma^2$, 
	and that $E_\delta(f_{i+1},\mathbf{x})$ is deterministic;
	and the last line follows from the same facts (primarily independence) plus that $\mathbb{E}[Z_j] = [0]$ for all $j$.
	(3) By independence and mean zero entries of $\Delta_i$,
	$\mathbb{E}\left[\langle A_i,B_i\rangle_F\right] ~=~0$ 
	since $A_i$ is independent of $\Delta_i$ while $B_i$ is linear in $\Delta_i$.
	(4) Finally, 
	$\|O(\delta)\|_F^2 = O(\delta^2)$, and cross terms with $O(\delta)$ are also $O(\delta^2)$ in expectation.

	Putting it all together gives
	\begin{align}
		\mathbb{E}\left[\|Z_i\|_F^2\right]
		\leq
		( \| E_\delta(h_i,\mathbf{z}_i) \|_2^2 + \text{dim}_2(\Delta_i) \sigma^2 )
		\cdot
		\mathbb{E}\!\left[\|Z_{i+1}\|_F^2\right]
		+
		\text{dim}_2(\Delta_i)  \sigma^2 \|E_\delta(f_{i+1},\mathbf{x})\|_F^2
		+
		O(\delta^2).
	\end{align}
	Therefore, assuming $\| E_\delta(h_i,\mathbf{z}_i) \|_2^2 \leq C_1$ 
	and $\| E_\delta(f_{i+1},\mathbf{x}) \|_F^2 \leq C_2$ for all $i$, 
	\begin{align*}
		\mathbb{E}\left[\|Z_i\|_F^2\right]
		&\leq
		(C_1 + \text{dim}_2(\Delta_i) \sigma^2)
		\cdot
		\mathbb{E}\left[\|Z_{i+1}\|_F^2\right]
		+  C_2 \text{dim}_2(\Delta_i) \sigma^2
		+ O(\delta^2) .
	\end{align*}
	Unrolling this recursion over $d - i$ layers, we get
	\begin{align*}
		\mathbb{E} \left[ \|\hat{E}_\delta(f_i, \bx) - E_\delta(f_i, \bx)\|_F^2 \right]
		&\leq
		\left( \prod_{\tau = i}^d (C_1 + \text{dim}_2(\Delta_\tau) \sigma^2) \right)
		\mathbb{E} \left[ \|\hat{E}_\delta(f_d, \bx) - E_\delta(f_d, \bx)\|_F^2 \right]
		\\
		&\hspace{30pt}
		+
		(C_2 \text{dim}_2(\Delta_i)  \sigma^2
		+ O(\delta^2)) 
			\sum_{\tau = i}^{d - 1} (C_1 + \text{dim}_2(\Delta_\tau) \sigma^2)^{\tau - i} .
	\end{align*}
	Although the statement above is more general than the one in the theorem statement, 
	we now simplify it for clarity of the result. 
	Under the assumption that $\text{dim}_2(\Delta_i) = \text{dim}_2(\Delta_j)$ for all $i, j \in [d]$,
	\begin{align*}
		\mathbb{E} \left[ \|\hat{E}_\delta(f_i, \bx) - E_\delta(f_i, \bx)\|_F^2 \right]
		&\leq
		(C_1 + \text{dim}_2(\Delta_i) \sigma^2)^{d-i} \mathbb{E} \left[ \|\hat{E}_\delta(f_d, \bx) - E_\delta(f_d, \bx)\|_F^2 \right]
		\\
		&\hspace{30pt}
		+
		(C_2 \text{dim}_2(\Delta_i)  \sigma^2
		+ O(\delta^2)) 
			\sum_{\tau = i}^{d - 1} (C_1 + \text{dim}_2(\Delta_i) \sigma^2)^{\tau - i} ,
	\end{align*}
	which implies
	\begin{align*}
		\mathbb{E} \left[ \|\hat{E}_\delta(f_1, \bx) - E_\delta(f_1, \bx)\|_F^2 \right]
		&\leq
		C_3^{d-1} \mathbb{E} \left[ \|\hat{E}_\delta(f_d, \bx) - E_\delta(f_d, \bx)\|_F^2 \right]
		+
		C_4  
			\sum_{\tau = 1}^{d - 1} C_3^{\tau - 1} ,
	\end{align*}
	as $\delta \rightarrow 0$, 
	where $C_3$ and $C_4$ are constants that depend on $\dim_2(\Delta)$ and $\sigma^2$.

	\paragraph{The bound is tight.}
	We now show that the bound is tight. 
	First, note that the only sources of ``slack'' in the upper bound above
	were (i) the use of the inequality $
	\mathbb{E}\left[\|A_i\|_F^2\right]
	\leq
	\| E_\delta(h_i,\mathbf{z}_i)\|_2^2
	\cdot
	 \mathbb{E} \left[ \|Z_{i+1}\|_F^2\right]
	$,
	and (ii) the assumption that $\| E_\delta(h_i,\mathbf{z}_i) \|_2^2 \leq C_1$ 
	and $\| E_\delta(f_{i+1},\mathbf{x}) \|_F^2 \leq C_2$ for all $i$.
	
	Suppose that $h_i : \mathbb{R}^\rho \to \mathbb{R}^\rho$ for all $i$.
	Let $E_\delta(h_i,\mathbf{z}_i) = \sqrt{C_1} \cdot \mathbb{I}_\rho$ for all $i$.
	Similarly, let $E_\delta (f_{i+1}, \mathbf{x}) = \sqrt{C_2 / \text{dim}(\bx)} \cdot \mathbb{I}_\rho$ for all $i$.
	Then (i) and (ii) hold with equality,
	and the bound is tight.

	\end{proof}

	\section{Additional Results for Section \ref{sec:explanations}}\label{app:additional_explainability_results}

\begin{figure}[ht]
    \centering
    \begin{subfigure}[b]{0.32\textwidth}
        \centering
        \includegraphics[width=\textwidth]{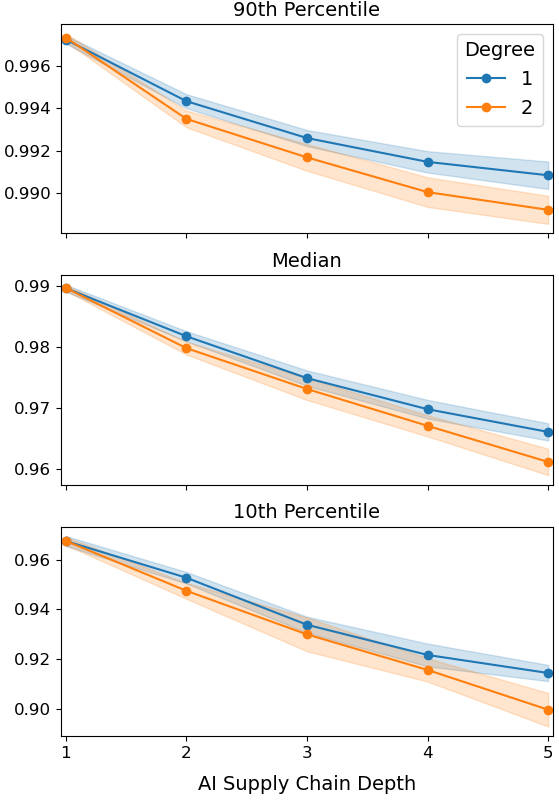}
        \vspace{-2pt}
        \caption{Cosine similarity}
        \label{fig:exp_cos_sim_2}
    \end{subfigure}
    \hfill
    \begin{subfigure}[b]{0.32\textwidth}
        \centering
        \includegraphics[width=\textwidth]{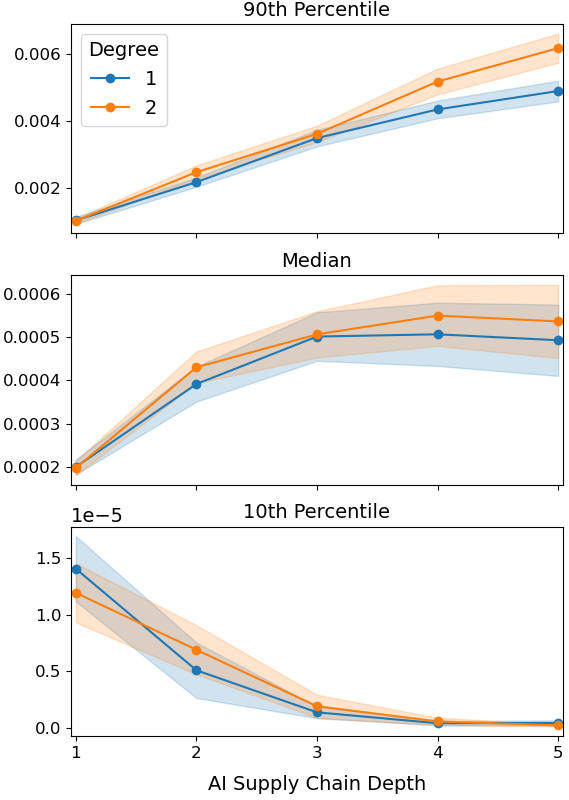}
        \vspace{-2pt}
        \caption{MSE}
        \label{fig:exp_mse_2}
    \end{subfigure}
    \hfill
    \begin{subfigure}[b]{0.32\textwidth}
        \centering
        \includegraphics[width=\textwidth]{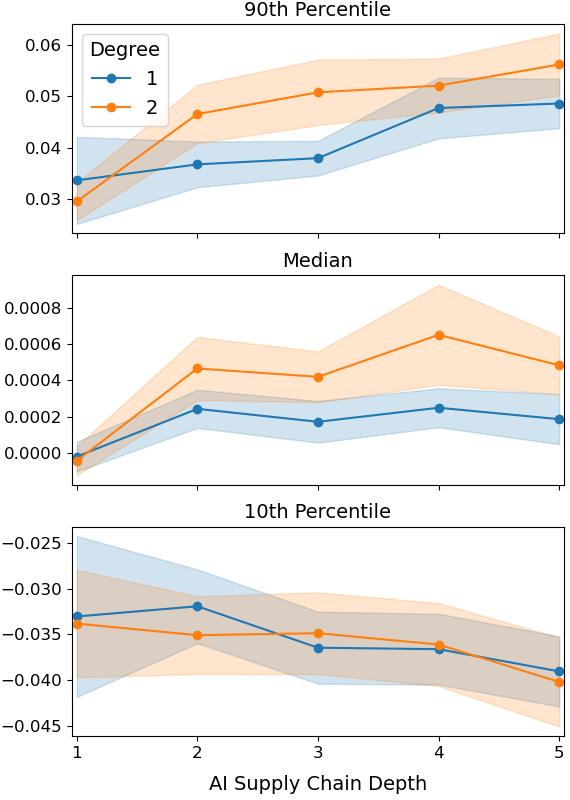}
        \vspace{-2pt}
        \caption{Recourse error}
        \label{fig:exp_rec_dist_2}
    \end{subfigure}
    \vspace{4pt}
    \caption{
        {
        Simulations for \Cref{sec:explanations_exp} for 50 LIME samples and a LIME radius of 0.1.
        }
    }
    \label{fig:explanations_additional_1}
\end{figure}

\begin{figure}[ht]
    \centering
    \begin{subfigure}[b]{0.32\textwidth}
        \centering
        \includegraphics[width=\textwidth]{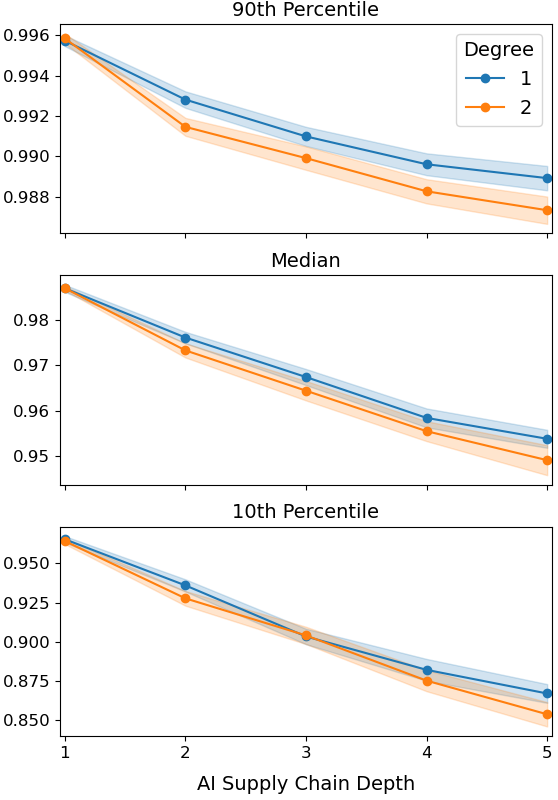}
        \vspace{-2pt}
        \caption{Cosine similarity}
        \label{fig:exp_cos_sim_3}
    \end{subfigure}
    \hfill
    \begin{subfigure}[b]{0.32\textwidth}
        \centering
        \includegraphics[width=\textwidth]{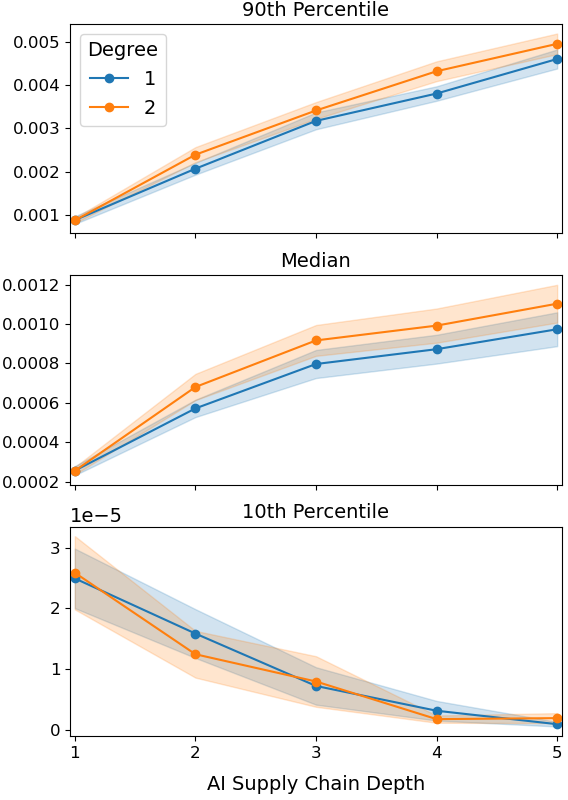}
        \vspace{-2pt}
        \caption{MSE}
        \label{fig:exp_mse_3}
    \end{subfigure}
    \hfill
    \begin{subfigure}[b]{0.32\textwidth}
        \centering
        \includegraphics[width=\textwidth]{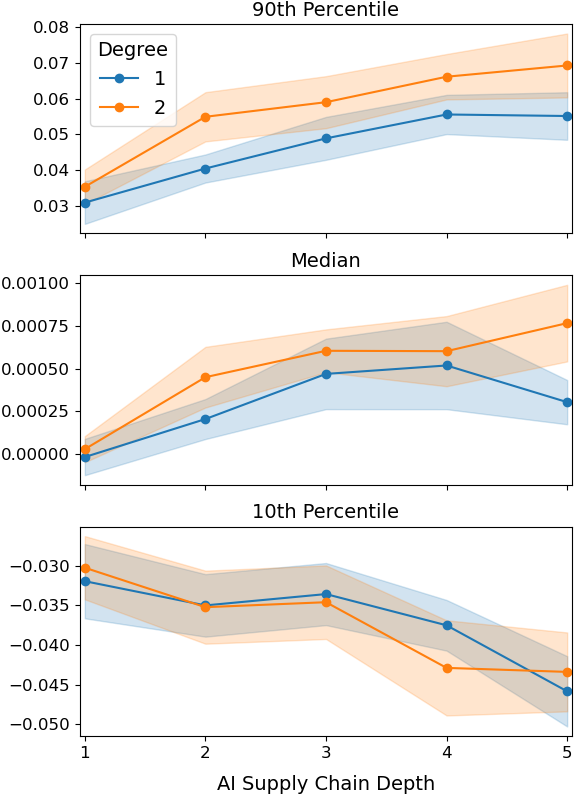}
        \vspace{-2pt}
        \caption{Recourse error}
        \label{fig:exp_rec_dist_3}
    \end{subfigure}
    \vspace{4pt}
    \caption{
        {
        Simulations for \Cref{sec:explanations_exp} for 100 LIME samples and a LIME radius of 0.2.
        }
    }
    \label{fig:explanations_additional_2}
\end{figure}

\begin{figure}[ht]
    \centering
    \begin{subfigure}[b]{0.32\textwidth}
        \centering
        \includegraphics[width=\textwidth]{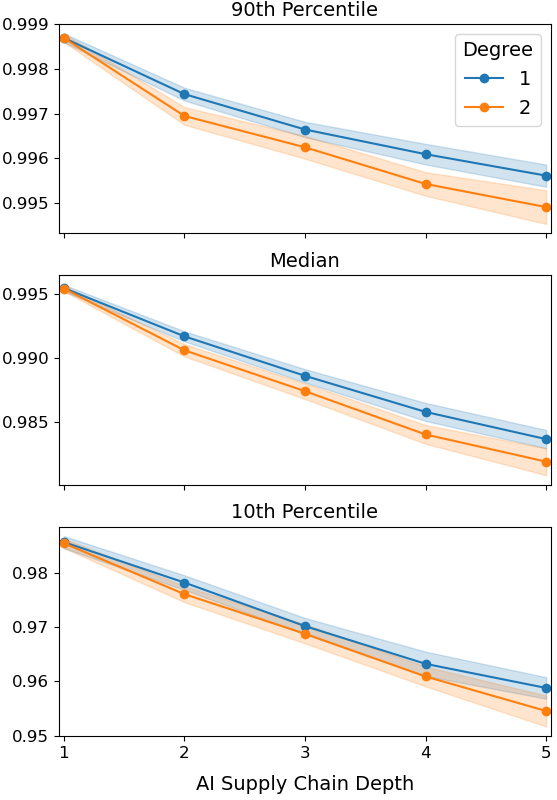}
        \vspace{-2pt}
        \caption{Cosine similarity}
        \label{fig:exp_cos_sim_4}
    \end{subfigure}
    \hfill
    \begin{subfigure}[b]{0.32\textwidth}
        \centering
        \includegraphics[width=\textwidth]{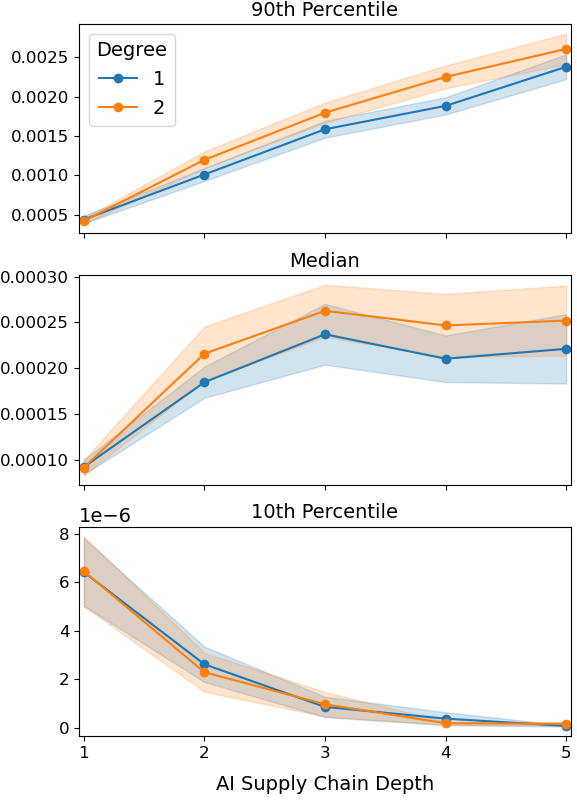}
        \vspace{-2pt}
        \caption{MSE}
        \label{fig:exp_mse_4}
    \end{subfigure}
    \hfill
    \begin{subfigure}[b]{0.32\textwidth}
        \centering
        \includegraphics[width=\textwidth]{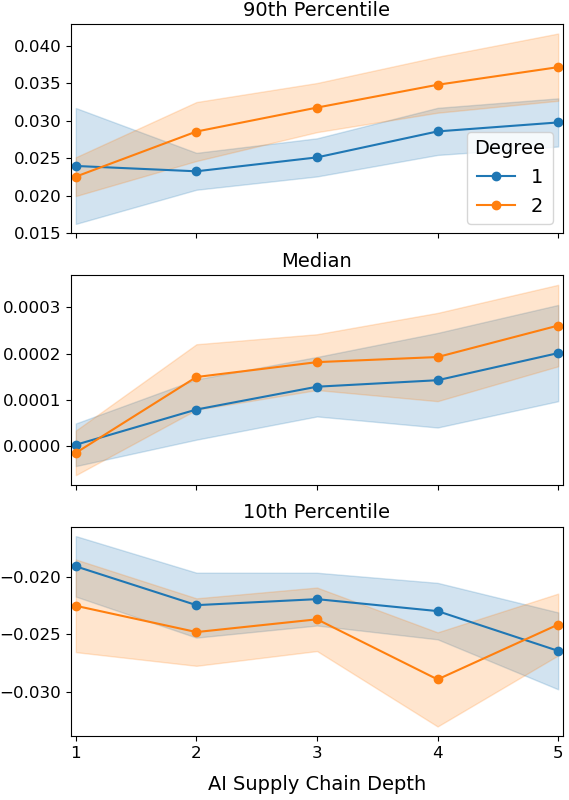}
        \vspace{-2pt}
        \caption{Recourse error}
        \label{fig:exp_rec_dist_4}
    \end{subfigure}
    \vspace{4pt}
    \caption{
        {
        Simulations for \Cref{sec:explanations_exp} for 100 LIME samples and a LIME radius of 0.1.
        }
    }
    \label{fig:explanations_additional_3}
\end{figure}

For \Cref{fig:explanations_additional_1,fig:explanations_additional_2,fig:explanations_additional_3},
the same plots as given in \Cref{sec:explanations_exp} are shown.
Specifically, from left to right, 
        we plot (a) the cosine similarity between end-to-end and supply-chain explanations,
        (b) the mean-squared error (MSE) between the two explanations,
        and (c) the recourse error between the two explanations.
        In all three figures (a)-(c),
        the x-axis is the depth of the ancestor tree and the legend gives the ``width'' or degree of the ancestor tree. 
        The three subplots show the means and 95\% confidence intervals of the 
        90th percentile, 50th percentile (median), and 10th percentile of the respective metric.

	\clearpage
\section{Proof of Theorem \ref{thm:fairness}}\label{sec:proof_fairness}

\begin{proof}
    Recall that the base model is trained such that $f_p(X)$ is approximately independent of $X_1$ given $Z$
    such that
    there exists $\varepsilon_p > 0$ for which
    \begin{align*}
            \left| \frac{\partial}{\partial x_1} \mathbb{E}_{D_p}[f_p(X) \mid X_1 = x_1, Z = z] \right| \leq \varepsilon_p, 
    \end{align*}
    for all $x_1 \in \mathbb{R}$ and $z \in \mathcal{Z}$.

    \paragraph{$f_p(X)$'s approximate conditional independence w.r.t. $X_1$ affects basis functions.}
    Using \Cref{eq:basis_function_f},
    we can write the approximate conditional independence condition as
    \begin{align*}
        \left| \frac{\partial}{\partial x_1} \mathbb{E} [ f_p(X) \mid X_1 = x_1, Z = z ] \right|
        &=
       \left| \frac{\partial}{\partial x_1} \mathbb{E} \left[ \sum_{i=1}^{N} \phi_i(X)  w_i \biggm| X_1 = x_1, Z = z \right] \right|
        =
        \left| \sum_{i=1}^{N} w_i g_i(x_1, z) \right| \leq \varepsilon_p,
    \end{align*}
    where $g_i(x_1, z) = \frac{\partial}{\partial x_1} \mathbb{E}_{D_p}  \left[ \phi_i(X) \mid X_1 = x_1, Z = z \right]$.
    This implies
    \begin{align}
        &\| \bw \| \| \mathbf{g}(x_1, z) \| | \cos (\bw, \mathbf{g}(x_1, z)) | \leq \varepsilon_p \nonumber
        \\
        &\implies
        \| \mathbf{g}_i(x_1, z) \|  \leq \frac{\varepsilon_p}{\| \bw \| | \cos (\bw, \mathbf{g}_i(x_1, z)) |} \label{eq:fp_basis_func} 
        \quad \text{ or } \quad
        | \cos (\bw, \mathbf{g}_i(x_1, z)) | = 0 ,
    \end{align}
    for all $x_1 \in \mathbb{R}$ and $z \in \mathcal{Z}$.
    
    \paragraph{$f_v(X)$'s approximate conditional independence w.r.t. $X_1$.}
    We now consider $f_v$. 
    Recall that $f_v$ learns a linear model on the embeddings (basis functions)
    that were learned by $f_p$.
    That is, optimizing $f_p$ involved learning both $w_i$ and $\phi_i$ for all $i$;
    optimizing $f_v$ involves learning $r_i$ for all $i$ for a fixed set of $\phi_i$'s. 

    Next, we examine what this means for the downstream fine-tuned model,
    where $r_i$'s are learned on the same basis functions $\phi_i$'s, 
    letting $\epsilon_v(x_1, z)$ denote the conditional independence ``tolerance''
    for $f_v$ at $(x_1, z)$:
    \begin{align}
        \varepsilon_v (x_1, z)
        :=&
        \left| \sum_{i = 1}^N r_i g_i(x_1, z) \right| \nonumber
        \\
        =& 
        \| \mathbf{r} \| \| \mathbf{g}_i(x_1, z) \| | \cos (\mathbf{r}, \mathbf{g}_i(x_1, z)) | \nonumber
        \\
        \leq& 
        \frac{\varepsilon_p \| \mathbf{r} \| | \cos(\mathbf{r}, \mathbf{g}_i(x_1, z))  | }{\| \bw \| | \cos (\bw, \mathbf{g}_i(x_1, z)) |} 
        \quad \forall x_1, z : \cos(\mathbf{w}, \mathbf{g}_i(x_1, z)) \neq 0,
        \label{eq:fv_cond_indep}
    \end{align}
    where the last line follows from \Cref{eq:fp_basis_func}.
    This establishes that approximate independence of $f_p(X)$ and $X_1$ given $Z$ 
    upstream does place an approximate conditional independence of $f_v(X)$ and $X_1$ given $Z$
    constraint downstream, but it is also approximate
    and may be looser (i.e., $\max_{x_1, z} \varepsilon_v(x_1, z)$ may be larger than $\varepsilon_p$)
    depending on the choice of $r_i$'s.
    Clearly, when $r_i$'s are chosen appropriately, 
    this threshold can be made arbitrarily large, 
    effectively removing the conditional independence constraint on $f_v$.

    Rearranging \Cref{eq:fv_cond_indep} implies
    \begin{align*}
        \| \mathbf{r} \|
        &\geq 
        \frac{\varepsilon_v (x_1, z) \| \bw \| | \cos (\bw, \mathbf{g}_i(x_1, z)) |}{
            \varepsilon_p   | \cos(\mathbf{r}, \mathbf{g}_i(x_1, z))  | } ,
        \qquad 
        \forall 
        x_1, z : \cos(\mathbf{w}, \mathbf{g}_i(x_1, z)) ,
            \cos(\mathbf{r}, \mathbf{g}_i(x_1, z)) \neq 0 .
    \end{align*}
    Letting $S = \{ (x_1, z) : 
        \cos(\mathbf{w}, \mathbf{g}_i(x_1, z)) \neq 0,
        \cos(\mathbf{r}, \mathbf{g}_i(x_1, z)) \neq 0 
         \}$,
    we can write the above as
    \begin{align*}
        \| \mathbf{r} \|
        &\geq 
        \max_{(x_1, z) \in S}
        \frac{\varepsilon_v (x_1, z) \| \bw \| | \cos (\bw, \mathbf{g}_i(x_1, z)) |}{
            \varepsilon_p   | \cos(\mathbf{r}, \mathbf{g}_i(x_1, z))  | } .
    \end{align*}
    This indicates that if $f_v$ wishes to \emph{loosen} the 
    type of conditional independence $f_p$ imposes
    (i.e., by increasing $\varepsilon_v(x_1, z)$),
    then doing so has implications on the lower bound of $\| \mathbf{r} \|$.

    \paragraph{Design choice of $f_v$'s developer.}
    So far, 
    we have not used the fact that $f_v$'s developer has their own (approximate) conditional independence constraint.

    However, as we see next, 
    the second type of conditional independence that the second developer seeks to satisfy 
    places an upper bound on $\| \mathbf{r} \|$.
    To see this, recall that $f_v$ is intended to satisfy a different notion of conditional independence,
    but $\phi$'s are already trained and unchangeable downstream;
    only $\mathbf{r}$ is changeable. 

    Let $g'_i(x_2, z) = \frac{\partial}{\partial x_2} \mathbb{E}_{D_v}  \left[ \phi_i(X) \mid X_2 = x_2, Z = z \right]$.
    Using the same flavor of analysis as applied to $f_p$ in \eqref{eq:fp_basis_func},
    we can write the conditional independence constraint on $f_v(X)$ and $X_2$ given $Z$ as
    \begin{align*}
        \left| \sum_{i = 1}^N r_i g'_i(x_1, z) \right| 
        &=
        \| \mathbf{r} \|
        \| \mathbf{g}'_i(x_2, z) \|
        | \cos (\mathbf{r}, \mathbf{g}'_i(x_2, z)) |
        \leq 
        \eta_v .
        \\
        \implies
        \| \mathbf{r} \| 
        &\leq \frac{\eta_v}{
            \| \mathbf{g}'_i(x_2, z) \|
        | \cos (\mathbf{r}, \mathbf{g}'_i(x_2, z)) | ,
        }
        \quad \text{ or } \quad
        | \cos (\mathbf{r}, \mathbf{g}'_i(x_2, z)) | = 0 , 
    \end{align*}
    for all $x_2 \in \mathbb{R}$ and $z \in \mathcal{Z}$.
    Let $S' = \{ (x_2, z) : 
        \cos(\mathbf{r}, \mathbf{g}'_i(x_2, z)) \neq 0 
         \}$.
    Therefore, $\| \mathbf{r} \|$  is bounded on both sides:
    \begin{align*}
        \max_{(x_1, z) \in S}
        \frac{\varepsilon_v (x_1, z ) \| \bw \| | \cos (\bw, \mathbf{g}_i(x_1, z)) |}{
            \varepsilon_p   | \cos(\mathbf{r}, \mathbf{g}_i(x_1, z))  | }
        \leq
        \| \mathbf{r} \| 
        &\leq 
        \min_{(x_2, z) \in S'}
        \frac{\eta_v}{
            \| \mathbf{g}'_i(x_2, z) \|
        | \cos (\mathbf{r}, \mathbf{g}'_i(x_2, z)) |
        } .
    \end{align*}
    Importantly, 
    when $\eta_v$ and $\varepsilon_v$ are \emph{inversely} related 
    (as in the case when two desiderata are not necessarily compatible, 
    as is true for many fairness constraints),
    then improving $f_v$'s ability to satisfy conditional independence of $f_v(X)$ and $X_2$ given $Z$
    (which is the second developer's goal) becomes challenging, 
    and they may in fact reach a point at which they can no longer
    decrease $\eta_v$ without violating the lower bound on $\| \mathbf{r} \|$. 
    In the final result, 
    we let $\eta_v = \max_{(x_1, z) \in S} \varepsilon_v (x_1, z) | \cos (\mathbf{w, g}_i(x_1, z)) |$.
\end{proof}

	\clearpage
\section{Additional Results for Section \ref{sec:fairness}}\label{app:additional_fairness_results}

\begin{figure}[ht]
    \centering
    \begin{subfigure}[b]{0.45\textwidth}
        \includegraphics[width=\textwidth]{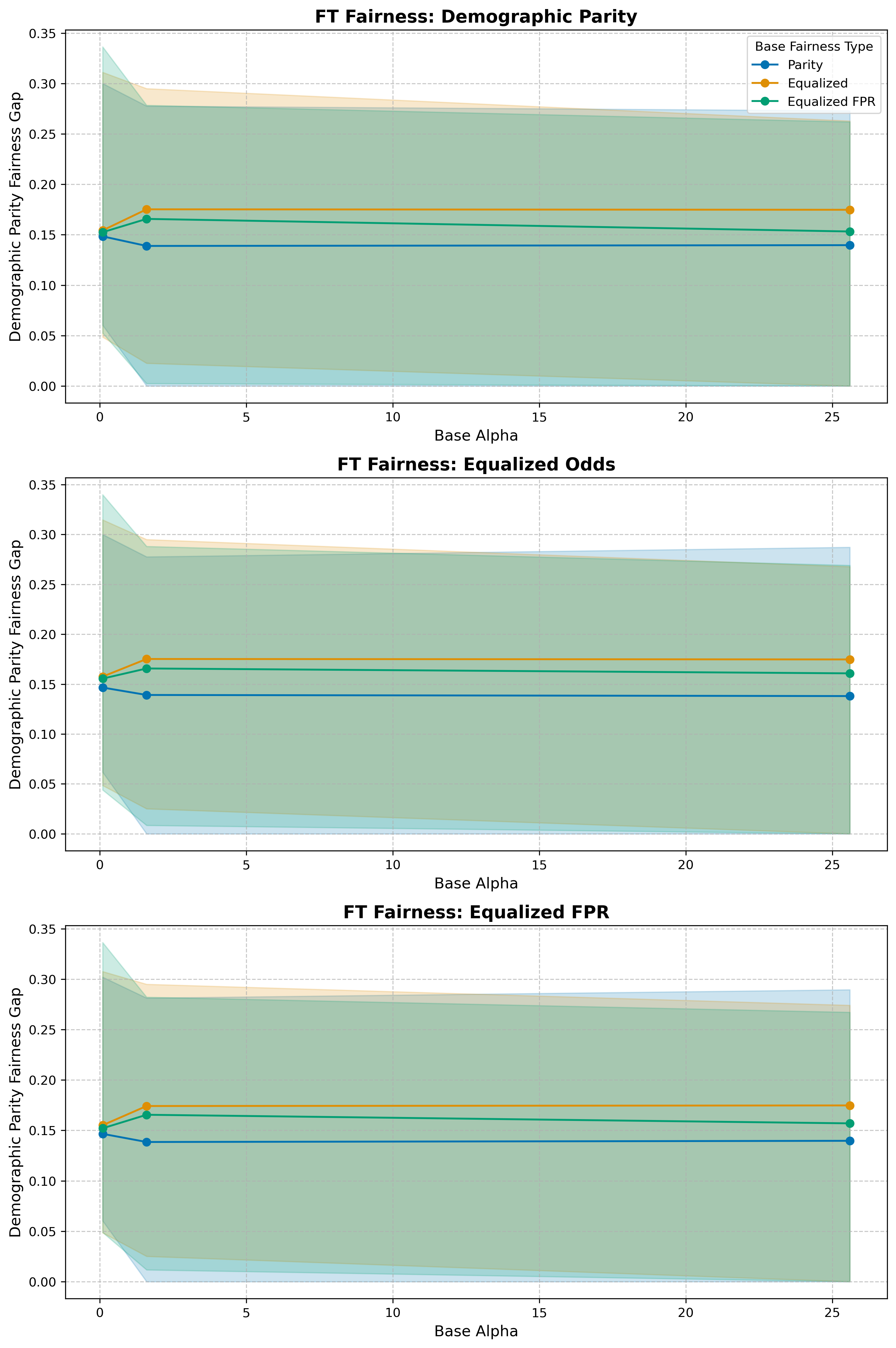}
        \caption{Demographic parity fairness gap}
        \label{fig:sub1}
    \end{subfigure}
    \hfill
    \begin{subfigure}[b]{0.45\textwidth}
        \includegraphics[width=\textwidth]{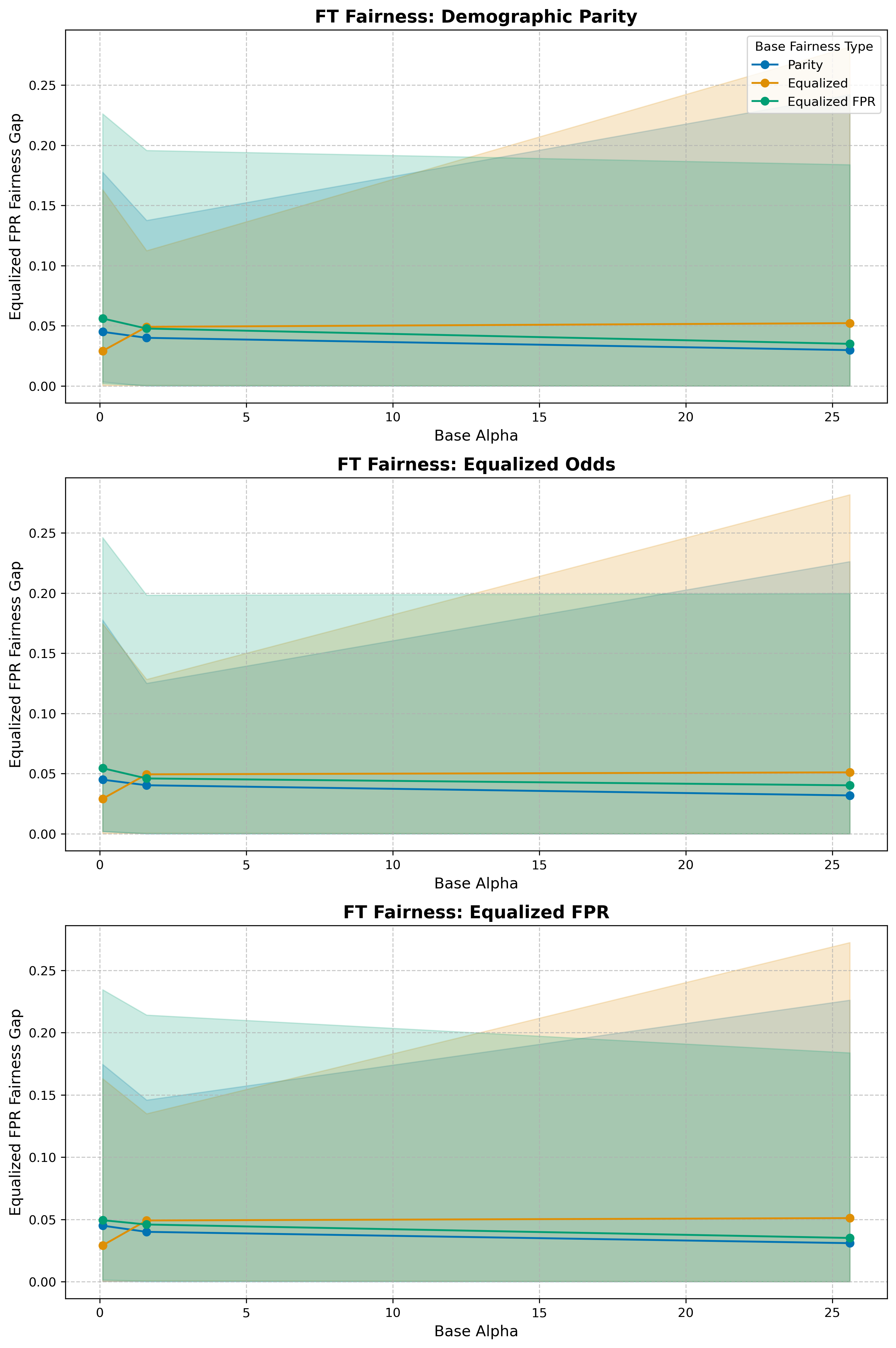}
        \caption{Equalized FPR fairness gap}
        \label{fig:sub2}
    \end{subfigure}
    \caption{The median (given by the points and line) 
    as well as maximum and minimum values (given by the shaded region)
    of the fairness gap across all downstream (fine-tuned) models. 
    In (\ref{fig:sub1}), the fairness gap on the y-axis is for demographic parity
    (the absolute difference between the positive rates of the two sensitive groups), 
    and in (\ref{fig:sub2}), the fairness gap on the y-axis is for equalized FPR
    (the absolute difference between the FPRs of the two sensitive groups).
    The x-axis gives the base $\alpha_p$ value used in training the upstream model, 
    and the colors denote the type of fairness used to train the upstream model.
    }
    \label{fig:fairness_reversible_max_min_ovelap}
\end{figure}

To support the claim that our experiments align with previous findings
(and our theoretical result) that upstream choices are ``reversible'' in the downstream model
(e.g., that information that has been ``unlearned'' upstream can be resurfaced by ``jogging the memory''
of the downstream model, or that fairness imposed upstream can be undone downstream), 
we add Figure \ref{fig:fairness_reversible_max_min_ovelap}, 
which plots the maximum, minimum, and median fairness gap values across our experiments.
Figure \ref{fig:fairness_reversible_max_min_ovelap}
shows that regardless of upstream choices, 
one can find downstream models that reverse the upstream choice 
(analogous to an existence result). 
This can be seen by the fact that the range of y-values overlap 
across colors (and across base $\alpha$ values) despite the different 
types of fairness used to train the upstream model,
i.e., there exists a way to fine-tune the downstream model such that the choice of upstream fairness
type and even the strength with which upsream fairness is enforced can be ``undone.''
As we discuss in the main text, 
results of this type seemingly suggest that upstream choices can be ``erased'' in downstream training, 
but as we show in Section \ref{sec:empirical_fairness}, 
an examination of multiple critera simultaneously reveals that upstream choices
affect downstream models by imposing additional trade-offs. 
That is, two facts can both be true:
(i) that upstream choices can be ``undone''
but (ii) that they still restrict downstream developers.

\end{document}